\documentclass[sigconf,edbt]{acmart-edbt2020}
\usepackage{hyperref}
\usepackage{ulem}
\usepackage{lineno}
\usepackage{url}
\usepackage{microtype}
\DeclareMathOperator*{\argmin}{argmin}
\usepackage{epsfig}
\usepackage{graphicx} 
\usepackage{epstopdf}
\usepackage{subfigure}
\usepackage{comment}
\usepackage{array}
\usepackage{pbox}
\usepackage{multirow}
\usepackage{enumitem}
\usepackage{todonotes}
\usepackage{caption}
\usepackage{mathtools}
\usepackage{breqn}
\usepackage{amsthm}
\usepackage{algorithm}
\usepackage[noend]{algpseudocode}
\usepackage{tikz-cd}
\DeclareGraphicsExtensions{.eps}

\algnewcommand\algorithmicforeach{\textbf{for each}}
\algdef{S}[FOR]{ForEach}[1]{\algorithmicforeach\ #1\ \algorithmicdo}

\begin{document}


\title{Finding Representative Sampling Subsets in Sensor Graphs using Time Series Similarities}

\author{Roshni Chakraborty}
\affiliation{%
  \institution{Aalborg University}
  \city{Aalborg}
  \country{Denmark}}
\email{roshnic@cs.aau.dk}

\author{Josefine Holm}
\affiliation{%
  \institution{Aalborg University}
  \city{Aalborg}
  \country{Denmark}}
\email{jho@es.aau.dk}

\author{Torben Bach Pedersen}
\affiliation{%
  \institution{Aalborg University}
  \city{Aalborg}
  \country{Denmark}}
\email{tbp@cs.aau.dk}

\author{Petar Popovski}
\affiliation{%
  \institution{Aalborg University}
  \city{Aalborg}
  \country{Denmark}}
\email{petarp@es.aau.dk}




\begin{abstract}
With the increasing use of IoT-enabled sensors, it is important to have effective methods for querying the sensors. For example, in a dense network of battery-driven temperature sensors, it is often possible to query (sample) just a subset of the sensors at any given time, since the values of the non-sampled sensors can be estimated from the sampled values. If we can divide the set of sensors into disjoint so-called \textit{representative sampling subsets} that each represent the other sensors sufficiently well, we can alternate the sampling between the sampling subsets and thus, increase battery life significantly. In this paper, we formulate the problem of finding representative sampling subsets as a graph problem on a so-called \textit{sensor graph} with the sensors as nodes. Our proposed solution, \textit{SubGraphSample}, consists of two phases. In Phase-I, we create edges in the sensor graph based on the similarities between the time series of sensor values, analyzing six different techniques based on proven time series similarity metrics. In Phase-II, we propose two new techniques and extend four existing ones to find the maximal number of \textit{representative sampling subsets}.  Finally, we propose \textit{AutoSubGraphSample} which auto-selects the best technique for Phase-I and Phase-II for a given dataset. Our extensive experimental evaluation shows that our approach can yield significant battery life improvements within realistic error bounds. 
\end{abstract}


\maketitle

 
 \section{Introduction}
\label{intro}
\par Recently, Internet of Things (IoT) enabled sensors are being widely used for different applications, such as, military operations, traffic management, home-service, healthcare, and several others~\cite{ashraf2020sagacious,chen2018query}. Irrespective of the application, the sensors generates continuous data, mostly in the form of time-series. This massive production of data has led to new challenges in data processing, storage and analysis ~\cite{paparrizos2021vergedb}. In order to handle this massive information overload, there is a need to develop effective methods that can query the sensors efficiently, for example, to preserve battery life. Identifying disjoint \textit{representative sampling subsets} such that only a representative sampling subset of sensors is queried at a given time. This is possible if the time-series generated by different sensors are similar. Therefore, a representative sampling subset represent the values of all the sensors sufficiently well~\cite{mao2016selection}. Therefore, in this paper, we aim to identify the maximum number of disjoint \textit{representative sampling subsets} on the basis of time-series similarity given the sensors and their time-series data.

\begin{figure}[ht]
\centerline{
\includegraphics[width=2in]{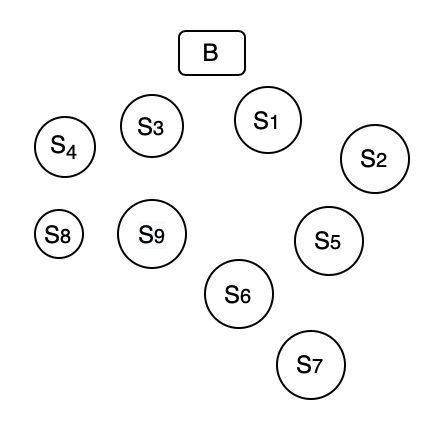}}
\caption{An overview of $SwN$ that comprises of sensors, $S_1$ to $S_{9}$, where each of the sensor records data in time-series, $T$. The sensors communicates to the base-station $B$ for processing and computation. The data recorded by sensors, $S$ is shown in Table \ref{tab:sgnres4}}
\label{fig:rep1}
\end{figure}

\par
\begin{figure}[ht]
\centerline{
\includegraphics[width=3.5in]{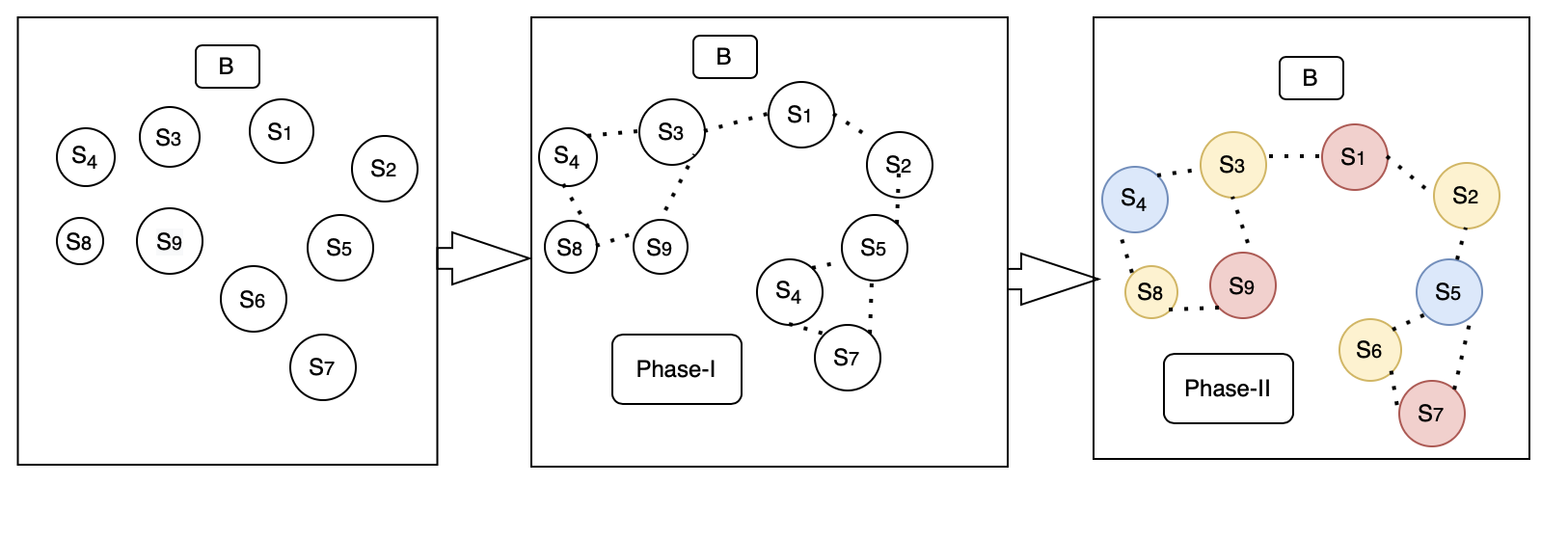}}
\caption{An overview of the Two-phase proposed framework, \textit{SubGraphSample} on $SwN$ is shown. In Phase I, we create a similarity graph, $G$ of the sensors where there is an edge between a pair of sensors if the similarity is greater than the threshold. In Phase II, we identify different \textit{representative sampling subsets} (represented by different colors) from $G$.}
\label{fig:rep2}
\end{figure}

 We discuss this through a motivating example now. For example, as shown in Figure \ref{fig:rep1}, $SwN$ comprises of $9$ sensors as  $\mathcal{S}={S_1, S_2, \cdots,S_9}$. Each of these sensors, $S_i$ records the temperature of a particular location as time-series, $T_i$ with $5$ time instances ${1, 2, \cdots,5}$ resulting in ${S_{i1}, S_{i2}, \cdots,S_{i5}}$ for a sensor, $S_i$. We simulate the data recorded by $\mathcal{S}$ similar to existing datasets \cite{diamond2013us}. We now compute the \textit{Fast DTW}~\cite{salvador2007toward} distance between the sensors (see Table \ref{tab:sgnres5}). The similarity is inversely proportional to the distance. For example, $S_1$ and $S_2$ are similar with a low distance of $8$ whereas $S_1$ and $S_9$, $S_2$ and $S_4$ are dissimilar with a high distance of $23$ and $28$, respectively. Therefore, we can query $S_1$ and $S_2$ at alternating timestamps, improving battery life while still getting sufficiently accurate results. Extending this idea to the entire sensor graph of $SwN$, we identify $3$ disjoint \textit{representative sampling subsets}, which each represent all the sensors within a given error bound on the time-series values. We provide an intuition in Figure \ref{fig:rep2} that by identifying $3$ disjoint \textit{representative sampling subsets} from $SwN$, we can increase the battery longevity by $3$ times compared to the case of querying all sensors at any given time. It is therefore required to devise a system  that can perform $a$) creation of a similarity graph of the sensors on the basis of the time-series data and, $b$) identification of the maximum number of \textit{representative sampling subsets} from the similarity graph.

\begin{table}{}
\centering
\begin{tabular}{|c|c|c|c|c|c|}
\hline
Sensor&$T$&Sensor&$T$\\
\hline\hline
$S_1$&$[4,5,5,5,4]$&$S_6$&$[7,9,10,10,9]$\\
\hline			h
$S_2$&$[6,6,7,7,5]$&$S_7$&$[9,9,9,11,7]$\\
\hline
$S_3$&$[1,1,3,3,3]$&$S_8$&$[0,3,3,3,0]$\\
\hline
$S_4$&$[0,0,1,1,1]$&$S_9$&$[1,1,1,4,1]$\\
\hline
$S_5$&$[8,8,8,8,6]$& &\\
\hline
\end{tabular}
\caption{The data recorded by $SwN$ which comprises of $9$ sensors that records temperature in $5$ time-instances}\label{tab:sgnres4}
\end{table}

\begin{table}{}
\centering
\begin{tabular}{|c|c|c|c|c|c|c|c|c|}
\hline
$S_i$&$S_j$&$T$&$S_i$&$S_j$&$T$&$S_i$&$S_j$&$T$\\
\hline\hline
$S_1$&$S_2$&$8$&$S_1$&$S_6$&$16$&$S_1$&$S_9$&$23$\\
\hline			
$S_2$&$S_5$&$7$&$S_2$&$S_4$&$28$&$S_2$&$S_7$&$16$\\
\hline
$S_3$&$S_1$&$12$&$S_3$&$S_2$&$20$&$S_3$&$S_8$&$5$\\
\hline
$S_6$&$S_5$&$9$&$S_6$&$S_7$&$6$&$S_6$&$S_8$&$36$\\
\hline
$S_9$&$S_8$&$7$&$S_9$&$S_5$&$30$&$S_9$&$S_3$&$4$\\
\hline
\end{tabular}
\caption{The distance between selected pair of sensors, $S_i$ and $S_j$ of $SwN$ by \textit{Fast DTW}~\cite{salvador2007toward}.}\label{tab:sgnres5}
\end{table}

\par Several selection sampling techniques in graph signal processing domain~\cite{tanaka2020sampling} have been proposed, including randomized~\cite{tanaka2020sampling,perraudin2018global} or deterministic greedy sampling~\cite{chamon2017greedy,gadde2014active} techniques which focus on finding a single \textit{representative sampling subset}. However, these approaches identify only one sub-set of sensors and therefore, do not solve our objective. Furthermore, existing sampling approaches rely on the availability of the graph topology of the sensors which might not be always available. Therefore, in this paper, we propose a Two-phase framework, namely, \textit{SubGraphSample}, that identifies the maximum number of \textit{representative sampling subsets} on the basis of similarity among the sensors such that sensors that generate similar data belong to different \textit{representative sampling subsets}. In  \textit{SubGraphSample}, we
initially create a similarity graph of the sensors in Phase-I and then, iteratively identify representatives from each possible subgraph of the similarity graph iteratively to form the maximal number of possible \textit{representative sampling subsets} in Phase-II.

\begin{enumerate}
    \item We propose \textit{SubGraphSample} that does not require the graph topology of the sensors and enable significant improvements in battery life. We compare $6$ similarity graph creation approaches in Phase-I, propose $2$ novel sampling techniques and extend $4$ existing sampling techniques in Phase-II. Our experimental evaluations on $4$ datasets show that the best combination of graph creation approach and sampling technique can provide $5-13$ times increase in battery life within a $20-40\%$ error bound given a dataset.
    
    \item We propose an auto-tuned algorithm, \textit{AutoSubGraphSample} to select the best possible combination of algorithm for Phase I and II given a dataset.
    
    \item Our evaluation of \textit{AutoSubGraphSample} on $4$ representative datasets shows that \textit{AutoSubGraphSample} can generalize well to new datasets.

\end{enumerate}    
    \par The organization of the paper is as follows. We discuss the existing research works in Section~\ref{s:rel} followed by the problem statement in Section~\ref{s:prob}. In Section~\ref{s:gtop} and~\ref{s:ggrp}, we discuss the proposed approach followed by the the experiments setup. We show our observations in Section~\ref{s:Exp} and finally, draw our conclusions and future works in Section~\ref{s:con}.

\section{Related Work}\label{s:rel}

\subsection{Phase-I: Creation of Similarity Graphs}
We categorize the existing research papers that attempt to create a similarity graph of sensors, given the data generated by the sensors, into three types of approaches: \textit{Statistical}, \textit{Time-Series Analysis} and \textit{Graph Signal Processing}.

\subsubsection{Statistical Approaches}
In order to identify graphs between sensors, a simple way is to calculate similarity between each pair of sensors and then, create an edge between them if their similarity is greater than the threshold~\cite{mateos2019connecting}. Therefore, existing metrics, such as the Pearson correlation, the Jaccard coefficient, the Gaussian radial basis function and mutual information are used to compute the pairwise similarity and thereby, identify the graph topology~\cite{egilmez2017graph,hassan2016topology}. Feizi et al.~\cite{feizi2013network} extended the pairwise correlation by including the indirect dependencies from the transitive correlations through a network deconvolution based approach.

\subsubsection{Time-Series Analysis based Approaches}
Using time-series data, one can compute the similarity between a pair of sensors and use it as a basis to define an edge in the graph. The existing approaches based on time-series can be classified into distance/neighbourhood based methods and feature based methods~\cite{jiang2020time}. Distance based methods focus on identifying different distance metrics to align a pair of time-series~\cite{wang2013experimental}. Traditional distance metrics that are inspired by the concept of edit distance~\cite{chen2004marriage} include Lp-norms~\cite{yi2000fast}, Euclidean Distance~\cite{faloutsos1994fast}, Dynamic Time Warping (DTW)~\cite{berndt1994using}, Longest Common Sub-sequence (LCSS)~\cite{vlachos2002discovering}, Edit Sequence on Real Sequence (EDR)~\cite{chen2005robust}, Swale~\cite{morse2007efficient}, Spatial Assembling Distance~\cite{chen2007spade}, etc. Further, several existing research papers have proposed different variants~\cite{cuturi2011fast} of these traditional distance metrics for different objectives, such as run-time~\cite{cuturi2017soft}, applicability to specific problem~\cite{yin2019new}, etc. Additionally, several recent research papers have proposed integration of both neighbourhood based metrics~\cite{jiang2020time,gong2018sequential} and distance based metrics to train machine learning models, such as, SVM, Random Forest and ensemble models~\cite{lines2015time}. Recently, several research papers have proposed different neural network architectures, autoencoders~\cite{abid2018autowarp}, deep networks~\cite{matsuo2021attention}, meta-learning based pre-training~\cite{narwariya2020meta}, attention modules~\cite{yao2020linear,matsuo2021attention} to capture the complex temporal relationships in time series.

\subsubsection{Graph Signal Processing based Approaches}
In order to ensure analysis and processing of the graph signals in both the vertex and the spectral domain of the graph, several recent papers infer an optimal graph topology such that the input data form graph signals with smooth variations on the resulting topology~\cite{venkitaraman2019predicting,liao2019learning}. Dong et al.~\cite{dong2015laplacian} propose a factor analysis based model which was extended by Kalofolias et al.~\cite{kalofolias2016learn} to include sparsity. 
However, these approaches assume smoothness of the graph signals used for training.

\subsubsection{Summary of Insights}
Considering the variety of existing approaches to infer the similarity graph based on the sensing data, there is still a lack of a study that compares how different approaches perform on a given dataset. In addition, to the best of our knowledge, there is no existing approach tailored to the application we are interested in. In this paper, we select several prominent existing approaches from the three categories described above and compare them in their role in Phase-I for a given dataset.

\subsection{Phase-II: Sampling Algorithms}
Randomized sampling based approaches~\cite{tanaka2020sampling,perraudin2018global,puy2018random} select nodes from a predetermined probability distribution. They have a low computational cost, but cannot ensure the same quality at each selection. Deterministic greedy sampling techniques resolve this by  selecting the optimal sensor at each iteration. This deterministic operation scales with polynomial complexity~\cite{chamon2017greedy,gadde2014active}. However, most of these sampling techniques search for only one optimal sampling set and do not consider the time dimension of the data \cite{kim2020qr,bai2020fast,sakiyama2019eigendecomposition}. Therefore, these techniques do not resolve our objective of maximizing battery longevity. The works \cite{ortiz2018sampling,wei2019optimal1} identify each sampling set representing a time-graph signal; nevertheless, the same node may participate in different \textit{representative sampling subsets} which is not suitable to maximize the battery longevity. The sampling technique from~\cite{chiumento2019energy} can ensure improvement in battery lifetime. However, \cite{holm2021lifetime} has shown that the approach from~\cite{chiumento2019energy} is suboptimal. In addition, several existing sampling techniques could be adjusted to identify multiple \textit{representative sampling subsets}. In this paper, we extend the sampling techniques from \cite{chamon2017greedy,chen2015discrete,tsitsvero2016signals} to find multiple \textit{representative sampling subsets}. Thus, we propose two novel sampling techniques and four variants of the existing sampling techniques to identify the maximum number of \textit{representative sampling subsets}.

\section{Problem Statement and Framework}\label{s:prob}

\subsection{Problem Statement}\label{s:prob1}
Given a sensor graph that comprise of $n$ IOT-enabled sensors, $\mathcal{S}=(S_1, S_2, \ldots, S_n)$, the time-series data for the sensor $S_i$ is denoted by $T_i$. Let $\mathcal{CP}$ denote the set of all (say, $q$ in this case) possible complete partitions of the network  $\mathcal{CP}=(\mathcal{SP}_1,\mathcal{SP}_2,\ldots,\mathcal{SP}_q)$. A partition, $\mathcal{SP}_u$ consists of several non-empty subsets of $\mathcal{S}$, i.e., $\mathcal{SP}_u=(SP_{u1},SP_{u2},\ldots,SP_{uk})$ such that $\bigcup\limits_{i=1}^{k} SP_{ui}=\mathcal{S}$. 
The sensors are battery-powered and have low computing power. We also assume that the time series data have no missing values. We intend to identify the optimal partition, $OSP$ from all the possible complete partitions, $\mathcal{CP}$ such that,

\begin{equation}
    \begin{aligned}
&OSP= \textrm{argmax}_{x \in \mathcal{CP}}(\lvert \mathcal{SP}_x \rvert) \\
s.t. \quad &Error( \mathcal{SP}_{xi}, \mathcal{S}) \leq \epsilon \quad \forall ~i \in \mathcal{SP}_x \\
&\mathcal{SP}_{xi} \cap \mathcal{SP}_{xj}=\emptyset \quad \forall ~(i\neq j) \in \mathcal{SP}_x 
%
\label{eq:probForm}
\end{aligned}
\end{equation}
The optimal partition, $OSP$ is the partition that comprises of the maximum number of \textit{representative sampling subsets}, $\mathcal{SP}_u$ such that each of these non-empty subsets, $\mathcal{SP}_{xi}$ can represent the values of all sensors well enough, i.e., the error in the information recorded by $\mathcal{SP}_{xi}$ when compared to $\mathcal{S}$ must be less than the threshold, $\epsilon$, as $(Error( \mathcal{SP}_{xi}, \mathcal{S}) \leq \epsilon)$. We consider reconstruction error to calculate $Error(\mathcal{SP}_{xi}, \mathcal{S})$ which we discuss in details in Section \ref{s:recon}. Additionally, we assume only periodic round robin scheduling of each \textit{representative sampling subset}, $\mathcal{SP}_{xi}$, of $OSP$ in this paper. Furthermore, we consider constraint that no two subsets of $OSP$ can overlap i.e., $\mathcal{SP}_{xi} \cap \mathcal{SP}_{xj}=\emptyset$. 

\par This problem can be reformulated to identify the optimal partition $OSP$ that minimizes the maximal error for a given number $K$ \textit{representative sampling subsets}:
\begin{equation}
\begin{aligned}
&OSP= argmin_{x \in \mathcal{CP}}(Error( \mathcal{SP}_{x}, \mathcal{S})) \\
s.t. &\lvert \mathcal{SP}_x \rvert=K \\
&\mathcal{SP}_{xi} \cap \mathcal{SP}_{xj}=\emptyset \quad \forall ~(i\neq j) \in \mathcal{SP}_x 
\label{eq:probForm1}
\end{aligned}    
\end{equation}

We propose a Two-phase framework, namely \textit{SubGraphSample} which can solve either of the two equivalent problems \eqref{eq:probForm} or \eqref{eq:probForm1}. In Phase-I, we create a similarity graph, $\mathcal{G}=(\mathcal{V},\mathcal{E})$ such that the vertices $\mathcal{V}$ are the sensors, $S$ and the edges $\mathcal{E}$ represent the similarity of the recorded data between each pair of sensors, $S_i$ and $S_j$ using existing approaches. In Phase-II, we identify the $OSP$ from $\mathcal{G}$. An overview of the proposed approach on $SwN$ is shown in Figure \ref{fig:rep2}. We discuss graph creation approaches for Phase-I in Section~\ref{s:gtop} and propose sampling approaches for Phase-II in Section~\ref{s:gtop}. Furthermore, we propose Algorithm \textit{AutoSubGraphSample} to recommend the most suitable algorithm for both Phase-I and Phase-II given a dataset. 

\subsection{Preliminaries}
We now discuss the graph signal processing preliminaries needed to understand the proposed sampling techniques and evaluation metrics. We consider a dataset which comprises of $n$ sensors, $t$ as the length of the time-series of each sensor such that $T = {0,1, \ldots ,(t-1)}$, $S={S_0, S_1, \ldots S_{n-1}}$ is the set of sensors and $s$ as signal for the rest of the paper.

\begin{itemize}
\item \textit{Degree Matrix, $D$:} A diagonal matrix that contains the degree of each node, i.e., with entries $D_{ii} = \sum_{j=1}^{N} A_{ij}$ and $D_{ij}$ = 0 for $i \neq j$, where $A$ is the adjacency matrix.
     \item \textit{Graph Laplacian, $L$:} $L$ is calculated as $L=A-D$, where $A$ is the adjacency matrix and $D$ is the degree matrix \cite{shuman2013emerging}.
    \item \textit{Signal:} A signal represents a time-dependent function that conveys information~\cite{ortega2018graph}. For example, the signal is $s_T={s_0,s_1, \ldots ,s_{t-1}}$ and $s_i$ is a sample of the signal $s_T$.

 \item \textit{Graph Signal, $x$:} A signal whose samples are indexed by the nodes of a graph \cite{ortega2018graph}. In this paper, we consider graph-time signal, i.e., one graph signal, $x^k$ per time stamp, $k$ where $k \in { {0,1, \ldots ,(t-1)}}$. Therefore, a graph signal represents the values of each sensor at a time-stamp, i.e., $x^k$ comprises of $n$ samples (for $n$ sensors) where each sample $x_j^k$ is the value for sensor, $S_j$ at the time stamp $k$. 
 
   \item \textit{Smoothness}: A graph signal $x^k$ at the $k-$th time-stamp is smooth if it has similar values for the neighbouring nodes of $\mathcal{G}$.

    \item \textit{Graph Fourier transform, $GFT$:} $GFT$ is the eigendecomposition of the graph Laplacian, $L$ or adjacency matrix, $A$ into eigenvalues, $\Lambda$ and eigenvectors, $V$. The eigendecomposition of $L$ is $L=V\Lambda V^{-1}$. $GFT$ of $x^k$, i.e., $\hat{x}^k$ which is defined as $\hat{x}^k=V^{-1} x^k$\cite{shuman2013emerging}.

    \item \textit{Bandlimited Signal:} This is a signal that is limited to have non-zero spectral density only for frequencies that are below a given frequency. If  $\hat{x}^k$ is bandlimited i.e. there exist a $\mathcal{B}\in\{0,1,...,F-1\}$ such that $\hat{x}_i=0$ for all $i\geq\mathcal{B}$, then $x^k$ is compressible and can be sampled \cite{ortega2018graph}.

    \item \textit{Singular Value Decomposition, $SVD$:} Singular value decomposition is a generalization of the eigenvalue decomposition, i.e., the the factorization of a matrix into a canonical form, whereby the matrix is represented in terms of its eigenvalues and eigenvectors. $SVD$ is $L=U\Sigma V^{-1}$, where $U$ is an $m \times m$ complex unitary matrix, $\mathbf{\Sigma}$ is an $m \times n$ rectangular diagonal matrix with non-negative real numbers on the diagonal, V is an $n \times n$ complex unitary matrix and if $U=V$, then it becomes an eigendecomposition \cite[Section 7.7]{friedbergelementary}.
\end{itemize}

\section{Phase I : Similarity Graph Creation}\label{s:gtop}
In this Section, we discuss the creation of the similarity graph,
$\mathcal{G}$, by selecting different approaches from Statistical Approaches,
Time Series based Approaches and Graph Signal Processing based
Approaches.

\subsection{Statistical Approaches}~\label{s:stat}
We discuss $2$ statistical approaches next.

\subsubsection{Correlation-based Approach, $P_{corr}$}
From~\cite{mateos2019connecting}, we calculate \textit{Pearson Correlation Coefficient}, $\mathcal{\rho}(S_i,S_j)$ to determine the similarity between $S_i$ and $S_j$ as:
\begin{eqnarray}
\mathcal{\rho}(S_i,S_j)= \frac{\sigma(S_i,S_j)} {\sqrt{var(S_i) \cdot var(S_j)}}\label{eq:Eq3}
\end{eqnarray}
\noindent where, $\sigma(S_i,S_j)$ is the co-variance between $S_i$ and $S_j$ and $var(S_i)$ calculates the variance of the data for $S_i$. Therefore, we create an edge 
between $S_i$ and $S_j$ in $G_{corr}$ if $\mathcal{\rho}(S_i,S_j)$ is greater than the threshold. 

\subsubsection{Network Deconvolution $P_{conv}$}
We use network deconvolution~\cite{feizi2013network,sulaimanov2016graph} to create $G_{conv}$ from the adjacency matrix $\mathcal{A}$. Network deconvolution calculates $\mathcal{A}$ based on the co-variance matrix, $\Sigma$, determined from the data $S$ generated by the sensors:
\begin{eqnarray}
\mathcal{A}= \Sigma(\mathbb{I}+\Sigma)^{-1}\label{eq:Eq4}
\end{eqnarray}

\subsection{Approaches based on Time-Series}~\label{s:time}
We discuss $3$ approaches that determine similarity based on the
time-series of each pair of sensors, $S_i$ and $S_j$.
\subsubsection{Dynamic Time Warping ($DTW$), $P_{dtw}$} \label{ss:DWT}
$P_{dtw}$ measures the distance between a pair of sensors, $S_i$ and $S_j$ by calculating the distance, $DisDTW(S_{i},S_{j})$ based on
the Euclidean distance of the respective time-series of the sensors, $S_{i}$ and $S_{j}$ at the particular time-stamp and the minimum of the cumulative distances of adjacent elements of the two-time series. However, $P_{dtw}$ incurs high computational cost which runs across different time-series. Therefore, we use \textit{Fast DTW}~\cite{salvador2007toward} which being an approximation of $P_{dtw}$ runs in linear time and space~\cite{salvador2007toward}. We calculate the distance between $S_i$ and $S_j$ as $DisFDTW(S_{i},S_{j})$ and create an edge between $S_{i}$ and $S_{j}$ in $G_{dtw}$ if the $DisFDTW(S_{i},S_{j})$ is less than the threshold.

\subsubsection{Edge Estimation based on Haar Wavelet Transform, $P_{haar}$} 
The data generated from the sensors is inherently unreliable and noisy. Therefore, we compress the time-series of sensor, $S_i$ to effectively handle the unreliability in the data by \textit{Haar wavelet transform}~\cite{chan2003haar}. We select the $K$-largest coefficient for $S_i$ and  $S_j$ to get a compressed approximation  as $S_i^'$ and $S_j^'$ respectively~\cite{wu2000comparison}. We create $G_{haar}$ in which an edge between $S_i$ and $S_j$ exists if the \textit{Euclidean distance} between  $S_i^'$ and $S_j^'$ is less than the threshold.

\subsubsection{K-NN Approach, $P_{nei}$}\label{s:nei}  
We follow \textit{K nearest neighbours}, where a class of a node is assigned on the basis of its $K$ nearest neighbours~\cite{altman1992introduction}. We initially calculate the distance between a pair of sensors, $S_i$ and $S_j$ based on \textit{Euclidean distance} and create an edge between $S_i$ and $S_j$ in $G_{nei}$ if the distance between them is among the least \textit{K-distances}. 

\subsection{Approaches based on Graph Signal Processing, $P_{gsp}$}\label{s:gsp}
We follow~\cite{kalofolias2016learn} to infer the graph topology from signals under the assumption that the signal observations from adjacent nodes in a graph form smooth graph signals. The solution from~\cite{kalofolias2016learn} is scalable and the pairwise distances of the data in matrix, $Z$, are introduced as in 
\begin{equation}
   W^\ast=\min_{W} \quad \lVert W \circ Z \rVert  + -\alpha\mathbf{1}^\top\log(W\mathbf{1})+\frac{\beta}{2}\Vert W\Vert_F^2
    \label{eq:Eq7}
\end{equation}
 where $Z$ is matrix with the data from the sensors with one row for each sonsor and one column for each time stamp, $W^\ast$ is the optimal weighted adjacency matrix, $\mathbf{1}^\top\log(W\mathbf{1})$ ensures overall connectivity of the graph by forcing the degrees to be positive while allowing sparsity, $\alpha$ and $\beta$ are parameters to control connectivity and sparsity respectively.
We follow the implementation in \cite{Pena17graph-learning} to determine the weighted adjacency matrix, $W$. We create an unweighted adjacency matrix, $A$ and graph, $G_{gsp}$ by creating an edge in $A$ and $G_{gsp}$ if the edge weight in $W$ is greater than threshold. However, we observe that most of the edge weights are around $0$ and very few edge weights are within $0.5-1$, therefore, it is difficult to create graphs with every edge density by $P_{gsp}$.

\subsection{Summary of Insights}\label{s:gsum}

In order to see how $P_{dtw}$ works, consider the following example. On the basis of distance calculated between each pair of sensors as shown in Table~\ref{tab:sgnres5}, we create an edge between each pair of sensors, $S_i$ and $S_j$ in $G_{dtw}$ if the distance between $S_i$ and $S_j$ is less than the threshold, say $15$ for $SwN$. Therefore, we show $G_{dtw}$ in Phase-I of the Figure~\ref{fig:rep2} where $S_1$ and $S_2$, $S_3$ and $S_9$ are connected as the distances are $8$ and $4$ which are less than $15$. Additionally, $S_1$ and $S_9$, $S_2$ and $S_4$ with distance $23$ and $28$ are not connected. In Section \ref{s:res}, we analyze the performance of each approach for Phase-I and then, provide dataset-based heuristics for selecting the best approach.

\section{Phase II: Identifying $OSP$}\label{s:ggrp}

We propose several sampling approaches that utilize $G$ to identify \textit{representative sampling subsets} are representative of the values of all sensors.

\subsection{Network Stratification based Approach, \textit{Strat}}\label{s:strat}

\par We propose a network stratification based sampling approach, \textit{Strat}, that captures the inter-relationship among sensors at group level to inherently handle the sparsity at individual connections and the generic global attributes at the network level\footnote{https://en.wikipedia.org/wiki/Level_of_analysis}.
Therefore, in \textit{Strat}, we initially group similar sensors together into communities by Modularity Maximization~\cite{blondel2008fast} followed by selecting representatives from each of these communities to create a \textit{representative sampling subset}. We use Modularity Maximization based community detection  to group similar sensors as it is similar to the problem of community detection in large networks, as in online social networks~\cite{leskovec2010empirical,leung2009towards}. Among the multiple available community detection algorithms, we have opted for Modularity Maximization as it is  efficient and scalable to large networks \cite{blondel2008fast}. In order to create a \textit{representative sampling subset}, we select a sensor from each community based on their importance to that community. We denote the importance a sensor, $S_i$ by \textit{NodeScore($S_i$)} and propose three different mechanisms to calculate \textit{NodeScore($S_i$)}. Therefore, we iteratively select sensors from each community in the decreasing order of \textit{NodeScore($S_i$)} to form a \textit{representative sampling subset}. We tune the selection method depending on whether we solve Equation \eqref{eq:probForm} or Equation \eqref{eq:probForm1}.

\begin{algorithm}					
		\caption{\textit{SRel} }\label{alg:SRel}
		\begin{flushleft}
			\hspace*{\algorithmicindent} \textbf{Input} \textit{ graph, $G=(V,E)$} and number of \textit{representative sampling subsets}, $K$\\
			
			\hspace*{\algorithmicindent} \textbf{Output} \textit{$OSP$}\\
		\end{flushleft}
		\begin{algorithmic}[1]
		\State $C$ = Modularity Maximization algorithm($G$)
		\State $ComMem= {\min_{\forall u \in C} Size(u)}/{{K}}$
		\For{$i$ in $S$}
			    \State \textit{NodeScore($S_i$)} = Eigenvector Centrality($S_i$) 
	     \EndFor
	     \State Initialize $SamplingSets$=[]
			    \For{$x$ in range(0,$K$)}
			    \For{$y$ in $C$}
			     \For{$z$ in \textit{ComMem}}
                    \State $MNodeScore={\max_{\forall i \in S} NodeScore(S_i)}$
                    \State Identify $S_i$ with $MNodeScore$
                    \State Add $S_i$ to $SamplingSets[x]$
             \EndFor        
			 \EndFor
			 \EndFor
			 \State $OSP$ = $SamplingSets$
			 \State Return $OSP$
\end{algorithmic}
\end{algorithm}

\subsubsection{Selection by Relevance, \textit{SRel}} \label{s:strat1}
In \textit{SRel}, we calculate \textit{NodeScore($S_i$)} as the relevance of $S_i$, $R_i$ with respect to $C_{k}$ to capture the ability of $S_i$ to represent all the sensors of a community, $C_{k}$. We measure $R_i$ by Eigenvector Centrality~\cite{ruhnau2000eigenvector}. We determine the number of sensors to be selected from $C_{k}$ by \textit{ComMem}. The calculation of \textit{ComMem} varies based on whether we solve the (\ref{eq:probForm}) or (\ref{eq:probForm1}). For (\ref{eq:probForm}), we select the minimum number of possible nodes from each community such that the selected nodes can represent all the nodes from the community sufficiently well. We set \textit{ComMem} to be $1$ and then, we iteratively select \textit{ComMem} sensors from $C_{k}$ in decreasing order of \textit{NodeScore($S_i$)} to create a \textit{representative sampling subset} such that the error of the \textit{representative sampling subset} with respect to all the sensors is less than $\epsilon$. We repeat these steps to create the maximum number of possible \textit{representative sampling subsets}, i.e., $OSP$ and thus, optimize (\ref{eq:probForm}). 


\par To solve (\ref{eq:probForm1}), we set \textit{ComMem} as the ratio of the size of the smallest community and the given number of \textit{representative sampling subset}s, $K$. We, then, iteratively select \textit{ComMem} sensors from a $C_{k}$ in decreasing order of \textit{NodeScore($S_i$)} to form a $\mathcal{SP}_{xi}$ and repeat this step for $K$ times to create $OSP$. We show the algorithm of \textit{SRel} in algorithm~\ref{alg:SRel}. We follow the same procedure as \textit{SRel} in \textit{SMMR} and \textit{SEMMR} to calculate \textit{ComMem} and determine $OSP$ for either  (\ref{eq:probForm}) or (\ref{eq:probForm1}). However, we calculate \textit{NodeScore($S_i$)} differently in \textit{SMMR} and \textit{SEMMR} which we discuss next.

\begin{algorithm}	
\caption{\textit{SMMR} }\label{alg:SMMR}
\begin{flushleft}
			\hspace*{\algorithmicindent} \textbf{Input} \textit{ graph, $G=(V,E)$} and number of \textit{representative sampling subsets}, $K$\\
			
			\hspace*{\algorithmicindent} \textbf{Output} \textit{$OSP$}\\
		\end{flushleft}
		\begin{algorithmic}[1]
		\State $C$ = Modularity Maximization algorithm($G$)
		\State $ComMem= {\min_{\forall u \in C} Size(u)}/{{K}} $
		\For{$i$ in $S$}
			    
			    \State \textit{NodeScore($S_i$)} = Eigenvector Centrality($S_i$) 
			  
	     \EndFor
	     \State Initialize sampling subsets, $SS$=[]
			    \For{$x$ in range(0,$K$)}
			    \For{$y$ in $C$}
			     \For{$z$ in \textit{ComMem}}
			        \State $A[S_i]$ = \textit{Adjacency List} of Node $S_i$
			         \State $D(a,b)$ = \textit{Difference} between a set $a$ and a set $b$
			        
			        \State Calculate $I_G(S_{i}, SS[x])=\lvert D(A[S_{i}],A[SS[x]])\rvert$ 
			        
                    \State Calculate $MNodeScore$ by Equation \ref{eq:valFunc}
                    \State Identify sensor, $S_i$ with $MNodeScore$
                    \State Add $S_i$ to $SamplingSets[x]$
                \EndFor
			     \EndFor
			     \EndFor
			  \State $OSP$ = $SamplingSets$
			 \State Return $OSP$
\end{algorithmic}
\end{algorithm}


\subsubsection{Selection by Maximum Marginal Relevance, \textit{SMMR}} \label{s:strat2}
In \textit{SMMR}, we consider both relevance and information gain of a sensor to calculate \textit{NodeScore($S_i$)}. We propose Maximum Marginal Relevance~\cite{carbonell1998use} based score to calculate \textit{NodeScore($S_i$)} which is the weighted average of the relevance, $R_i$ and the information gain provided by $S_i$ with respect to $\mathcal{SP}_{xi}$, $IG(S_i,\mathcal{SP}_{xi})$. We measure $R_i$ as in \textit{SRel} and $IG(S_i,\mathcal{SP}_{xi})$ as the difference between the adjacency list of $S_i$ and the adjacency list of the already selected sensors in $\mathcal{SP}_{xi}$. Therefore, we select the sensor with maximum node score, \textit{MNodeScore} and further, repeat this for \textit{ComMem} times for each $C_{k}$ iteratively. The calculation of \textit{MNodeScore} is as follows
\begin{align}
\textit{MNodeScore}&=\max_{\{S_{i} \in C_{k}\}} [NodeScore(S_{i})] \\
&=\max_{S_{i} \in C_{k}} [\beta  R_i - (1- \beta)I_G(S_{i}, \mathcal{SP}_{xi}] \quad \label{eq:valFunc}
\end{align}
where $\beta$ is the weight for relevance and $(1-\beta)$ for information gain respectively. For our experiments, we consider $\beta$ as $0.4$. We show the pseudocode of \textit{SMMR} in Algorithm~\ref{alg:SMMR}.

\subsubsection{Selection by Error based Maximum Marginal Relevance, \textit{SEMMR}} \label{s:strat3}
In \textit{SRel} and \textit{SMMR}, we consider the edges between the sensors in $G$ to calculate \textit{NodeScore($S_i$)} and do not consider the actual data generated by $S_i$. In \textit{SEMMR}, we incorporate this information by calculating $IG(S_i,\mathcal{SP}_{xi})$ as the average of the minimum square error between the data generated by $S_i$ and the other sensors already selected in $\mathcal{SP}_{xi}$. We follow the same procedure of \textit{SMMR} to calculate \textit{MNodeScore} and finally, follow the same procedure as discussed in \textit{SRel} to determine \textit{ComMem} and resolve either Equation (\ref{eq:probForm}) or Equation (\ref{eq:probForm1}) accordingly. Algorithm~\ref{alg:SEMMR} shows the pseudocode of \textit{SEMMR}.


\begin{algorithm}	
\caption{\textit{SEMMR} }\label{alg:SEMMR}
\begin{flushleft}
			\hspace*{\algorithmicindent} \textbf{Input} \textit{ graph, $G=(V,E)$} and number of \textit{representative sampling subsets}, $K$\\
			
			\hspace*{\algorithmicindent} \textbf{Output} \textit{$OSP$}\\
		\end{flushleft}
		\begin{algorithmic}[1]
		\State $C$ = Modularity Maximization algorithm($G$)
		\State $ComMem= {\min_{\forall u \in C} Size(u)}/{{K}} $
		\For{$i$ in $S$}
			    \State \textit{NodeScore($S_i$)} = Eigenvector Centrality($S_i$) 
	     \EndFor
	     \State Initialize $SamplingSets$=[]
			    \For{$x$ in range(0,$K$)}
			    \For{$y$ in $C$}
			     \For{$z$ in \textit{ComMem}}
			        \State $T[S_i]$ = \textit{Time-Series Data} of $S_i$
			         \State $Er(S_i,SS[x])$ = \textit{Average Minimum Square Error} of $S_i$ wrt $SS[x]$
			        \State Calculate $I_G(S_{i}, SS[x])= Er(T[S_{i}],T[SS[x]])$
                   \State Calculate \textit{MNodeScore} by Equation \ref{eq:valFunc}
                    \State Identify sensor, $S_i$ with \textit{MNodeScore}
                    \State Add $S_i$ to $SS[x]$
			     \EndFor
			     \EndFor
			     \EndFor
			 \State $OSP$ = $SamplingSets$
			 \State Return $OSP$
\end{algorithmic}
\end{algorithm}


\subsection{Minimum singular value based approach, \textit{MSV}}\label{sec:samp_chen}

Chen et al. \cite{chen2015discrete} proposed a greedy-selection based sampling algorithm in which they iteratively select the node that maximizes the minimum singular value of the eigen vector matrix under the assumption that the signal is bandlimited. Under the assumption of bandlimitedness, selection of the best $\vert\mathcal{B}\vert$ nodes ensures almost complete reconstruction of the graph signal given that there is no sampling noise. Therefore, by choosing the node that maximizes the minimum singular value, Chen et al. optimize the information in the graph Fourier domain and forms a greedy approximation of the best $\vert\mathcal{B}\vert$ nodes. The authors consider eigendecomposition of the adjacency matrix, $A=U\Sigma U^{-1}$, for graph Fourier transform, $\hat{x}=U^{-1}x$ and create only one \textit{representative sampling subset} with $\vert\mathcal{B}\vert$ nodes by selecting the nodes iteratively according to:
\begin{equation}\label{eq:chen}
    m=\text{argmax}_q\sigma_{min}(U_{\mathcal{B},A_p+\{q\} })
\end{equation}
where $U_{\mathcal{B},A}$ is the first $\vert\mathcal{B}\vert$ rows of $U$, $A$ represents the set of columns of $U$ and $\sigma_{min}(U)$ is the function for the minimal singular value of $U$. In this paper, we propose \textit{MSV} which is an extension of \cite{chen2015discrete} where we generate $K$ \textit{representative sampling subsets} by iteratively adding nodes to each \textit{representative sampling subset} according to Equation \eqref{eq:chen} until all nodes have been assigned. We provide the pseudocode of \textit{MSV} in Algorithm \ref{alg:chen}. Applying \textit{MSV} for $SwN$ to generate $3$ \textit{representative sampling subsets} results in (5,4,7), (2,6,3) and (0,8,1).


\begin{algorithm}
\caption{Minimum Singular Value, $MSV$}\label{alg:chen}
\begin{algorithmic}[1]
\State{Input: $U$, $\mathcal{B}$, $k$}
\State $OSP$ an Assembly of $k$ empty sets
\For{$i$ in the number of nodes}
\State $p=i\textbf{ mod }k$
\State $m=$argmax$_q\sigma_{min}(U_{\mathcal{B},OSP_p+\{q\} })$
\State $OSP_p\leftarrow OSP_p+\{m\}$
\EndFor
\State Return $OSP$
\end{algorithmic}
\end{algorithm}

\subsection{Greedy MSE Based Approach, \textit{JIP and SIP}} \label{sec:samp_sip}

We propose two sampling techniques, i.e., Joint Iterative Partitioning, \textit{JIP}, and Simultaneous Iterative Partitioning, \textit{SIP}, that consider \textit{Mean Square Error} to select a node into a \textit{representative sampling subset}. By considering \textit{Mean Square Error}, we ensure that each \textit{representative sampling subset} generated can reconstruct the original graph within an error bound. \textit{JIP} and \textit{SIP} differs on the basis of the problem they intend to solve. \textit{JIP} \cite{holm2021lifetime} identify the maximum number of possible \textit{representative sampling subsets} given the $MSE$ to solve Equation \eqref{eq:probForm} and \textit{SIP} minimizes the $MSE$ given the number of possible \textit{representative sampling subsets} to solve Equation \eqref{eq:probForm1}. We discuss \textit{JIP} and \textit{SIP} in details next. We estimate \textit{MSE} as in \cite{chamon2017greedy}:
\begin{equation}
    \text{MSE}(OSP_p)=Tr[Q(OSP_p)]
\end{equation}
where
\begin{equation}
    Q(OSP_p)=V_{\mathcal{B}}\left(\Lambda^{-1}+\sum_{i\in OSP_p}\eta_{i}^{-1}v_iv_i^H\right)^{-1}V_{\mathcal{B}}^H
\end{equation}
Here $OSP_p$ is the $p$'th \textit{representative sampling subset}, $V_{\mathcal{B}}$ is the first $\vert\mathcal{B}\vert$ columns of the eigenmatrix, $V$, $v_i$ is the $i$'th row of $V$ and $\eta_{i}$ is the $i$'th entry in $\eta$ which is the variance of the noise. 
Therefore, $MSE$ is calculated iteratively as nodes are added to a \textit{representative sampling subset}. The reformulation is thoroughly described in \cite{holm2021lifetime} as:
\begin{equation}\label{eq:MSE_add}
    \text{MSE}(OSP_{p,j}\cup v_s)=Tr[Q_{j}]-\frac{v_s^HQ_{j}V_{\mathcal{B}}^HV_{\mathcal{B}}Q_{j}v_s}{\eta_{s}+v_s^HQ_{j}v_s},
\end{equation}
where
\begin{equation}\label{eq:K_j}
    Q_{j}=Q_{j-1}-V_{\mathcal{B}}^HV_{\mathcal{B}}\frac{Q_{j-1}v_uv_u^HQ_{j-1}}{\eta_{u}+v_u^HQ_{j-1}v_u},
\end{equation}
$Q_0=\Lambda$ and $u$ is the index for the most recently added node.
\begin{equation}\label{eq:MSE_node}
    MSE(v_s)=MSE(\emptyset\cup v_s)
\end{equation}

In \textit{JIP}, we iteratively create \textit{representative sampling subsets} such that the $MSE$ of each \textit{representative sampling subsets} is within the $MSE$ threshold. Therefore, we initially create a \textit{representative sampling subset} by adding the node with the least $MSE$ according to Equation \eqref{eq:MSE_node} until the $MSE$ of that \textit{representative sampling subset} is less than the threshold. We, further, repeat this for the maximum number of possible \textit{representative sampling subsets}. If there are any nodes left that can not form a \textit{representative sampling subset} on their own, they are divided among existing \textit{representative sampling subsets}. The pseudocode of \textit{JIP} is shown in Algorithm \ref{Alg1}. 

\begin{algorithm}
\caption{Joint Iterative Partitioning, $JIP$}\label{Alg1}
\begin{algorithmic}[1]
\State{Input: $\Lambda,\eta, V, \epsilon$}
\State$Q_0=\Lambda, p=0,j=1$
\State $L_s=MSE(v_s)$ 
\State $L\_index=argsort(L)$ (largest first)
\While{$L\_index\neq \emptyset$}\label{while1}
\State $u=pop(L\_index_{-1})$
\State $OSP_{p,j}=u$
\While{$MSE(OSP_{p,j})>\epsilon$ and $L\_index\neq \emptyset$}
\State Calculate $Q_j$ according to \eqref{eq:K_j}
\For{$i$ in $L\_index$}
\If{$MSE(OSP_{p,j}\cup L_i)<\epsilon$}
\State $OSP_{p,j+1}=\{OSP_{p,j}\cup L_i\}$
\State $j=0,p=p+1,delete(L\_index=i)$
\State goto \ref{while1}
\EndIf
\EndFor
\State $OSP_{p,j+1}=\{OSP_{p,j}\cup \text{arg}\min(MSE(L))\}$
\State remove chosen node from $L\_index$, $j=j+1$
\EndWhile
\EndWhile
\If {$MSE(OSP_{-1,-1})>\epsilon$}
\State Split the nodes in $OSP_{-1}$ among the other sets
\EndIf
\State Return $OSP$
\end{algorithmic}
\end{algorithm}

In \textit{SIP}, we aim to generate \textit{representative sampling subsets} for better and more balanced solutions to Equation \eqref{eq:probForm1}. Given the number of \textit{representative sampling subsets}, $k$, at each iteration, \textit{SIP} creates the \textit{representative sampling subsets} simultaneously unlike \textit{JIP}. After sorting the nodes according to Equation \eqref{eq:MSE_node}, the $k$ nodes with the lowest $MSE$ are added to the \textit{representative sampling subsets}, such that each \textit{representative sampling subset} has been assigned one node. At each iteration, we add the best node according to \eqref{eq:MSE_add} to the \textit{representative sampling subset} with the largest $MSE$ and repeat this for all the \textit{representative sampling subsets} in the same order. We iterate this until all the nodes are allocated to a \textit{representative sampling subset}. The pseudocode of \textit{SIP} is given in Algorithm \ref{Alg2}. The \textit{representative sampling subsets} by \textit{JIP} are (5,4,7), (2,6,3) and (0,8,1) and by \textit{SIP}  are (5,1,3), (4,8,0) and (2,6,7) respectively.

\begin{algorithm}
\caption{Simultaneous Iterative Partitioning, $SIP$}\label{Alg2}
\begin{algorithmic}[1]
\State{Input: $K$}
\State list of k empty arrays, $OSP$
\State$Q=array(\Lambda,K),err=zeros(K)$
\State $L_s=MSE(v_s)$ 
\State $L\_index=[0,...,N]$
\For{$i$ in range$(K)$}
\State $m=\text{arg}\min(L_s)$
\State append $m$ to $OSP_i$
\State $err[i]=L_s[m]$
\State delete $L_s[m]$ and $L_{index}[m]$
\State Update $Q_i$ according to \eqref{eq:K_j}
\EndFor
\While{$L\_index\neq \emptyset$}\label{while2}
\State $j=\text{arg}\max(err)$
\State $m=ones(N)\max(L_s)$
\For{$i$ in $L\_index$}
\State $m[i]=MSE(OSP_j\cup v_i)$
\EndFor
\State append $\text{arg}\min(m)$ to $OSP_j$
\State $err[j]=\min(m)$
\State Update $Q_i$ according to \eqref{eq:K_j}
\State delete $l_{index}=m$
\EndWhile
\State Return $OSP,err$
\end{algorithmic}
\end{algorithm}

\subsection{Minimum Frobenius Norm, \textit{Frob}, and Maximum Parallelepiped Volume, \textit{Par}} \label{sec:samp_par}


Tsitsvero et al. \cite{tsitsvero2016signals} proposed two different greedy based sampling algorithms, $\textit{GFrob}$ and $\textit{GPar}$, 
which are based on eigendecomposition of graph Laplacian, $L=V\Sigma V^{-1}$, under the assumption that the signal is $\mathcal{B}$-bandlimited. $\textit{Frob}^'$ aims to find the \textit{representative sampling subset} of size $\vert\mathcal{B}\vert$ that minimizes the frobenius norm for the  pseudo-inverse for the eigenvector matrix restricted to the first $\vert\mathcal{B}\vert$ columns and the rows corresponding to the chosen \textit{representative sampling subset}, i.e.

\begin{equation}
    OSP_i=\argmin_{P_i\in S:\vert P_i\vert=\vert\mathcal{B}\vert}\Vert(V_{\mathcal{B},P_i})^+\Vert_F,
    \label{eq:frob}
\end{equation}
where $V_{\mathcal{B},P_i}$ is  and the columns of the set $P_i$ of $V$, $V^+$ denotes the  pseudo-inverse and $S$ is the set of all nodes. However, $\textit{GFrob}$ generates only one \textit{representative sampling subset} by adding $\vert\mathcal{B}\vert$ nodes in a greedy manner according to: 

\begin{equation}\label{eq:Frobenius}
    m=\argmin_q\sum_j\frac{1}{\sigma_j^2(V_{\mathcal{B},A_p+\{q\} })}.
\end{equation} 
where $\sigma_j^2(V)$ denotes the $j$'th singular value of $V$.

In this paper, we extend $\textit{GFrob}$ as \textit{Frob} to generate $k$ \textit{representative sampling subsets} by adding a node to a \textit{representative sampling subset} on the basis of Equation \eqref{eq:Frobenius} in a round robin manner until all the nodes have been assigned to a \textit{representative sampling subset}. An overview of \textit{Frob} is shown in Algorithm \ref{alg:Frobenius}.  Similarly, $\textit{Par}$, selects the nodes $\vert\mathcal{B}\vert$ nodes in a greedy manner according to:


\begin{algorithm}
\caption{Minimum Frobenius Norm, \textit{Frob}}\label{alg:Frobenius}
\begin{algorithmic}[1]
\State{Input: $U$, $\mathcal{B}$, $k$}
\State $OSP$ an Assembly of $k$ empty sets
\For{$i$ in the number of nodes}
\State $p=i\textbf{ mod }k$
\State $m=$argmin$_q\sum_j\frac{1}{\sigma_j^2(U_{\mathcal{B},OSP_p+\{q\} })}$
\State $OSP_p\leftarrow OSP_p+\{m\}$
\EndFor
\State Return $OSP$
\end{algorithmic}
\end{algorithm}

\begin{equation}\label{eq:Parallel}
    m=\argmin_q\prod_j\lambda_j(V_{\mathcal{B},A_p+\{q\} }V_{\mathcal{B},A_p+\{q\} }^H).
\end{equation}
where $\lambda_j(V)$ is the $j$'th eigenvalue of $V$.
For \textit{Par}, we follow the same approach as proposed in \textit{Frob} to identify the maximum number of possible \textit{representative sampling subsets} that optimizes Equation \eqref{eq:Parallel} instead of Equation \eqref{eq:Frobenius}. An overview of \textit{Par} is shown in Algorithm \ref{alg:Parallel}. The \textit{representative sampling subsets} generated by \textit{Frob} and \textit{Par} on $G_{dtw}$ for $SwN$ are same which are (8,2,7), (4,5,3) and (0,6,1) respectively.


\begin{algorithm}
\caption{Maximum Parallelepiped Volume, \textit{Par}}\label{alg:Parallel}
\begin{algorithmic}[1]
\State{Input: $V$, $\mathcal{B}$, $k$}
\State $A$ an Assembly of $k$ empty sets
\For{$i$ in the number of nodes}
\State $p=i\textbf{ mod }k$
\State $m=$argmin$_q\prod_j\lambda_j(V_{\mathcal{B},A_p+\{q\} }V_{\mathcal{B},A_p+\{q\} }^H)$
\State $A_p\leftarrow A_p+\{m\}$
\EndFor
\State Return $A$
\end{algorithmic}
\end{algorithm}


\subsection{AutoSubGraphSample}~\label{s:exp15} 
\par We have discussed $6$ existing graph creation approaches for Phase-I and proposed $6$ sampling techniques for Phase-II. However, the performance of these approaches differ across different datasets as they have different properties. Therefore, there is a need to automatically select the most suitable approach for Phase-I and Phase-II respectively given a dataset. In this Subsection, we propose an Algorithm \textit{AutoSubGraphSample} that considers the meta data of the dataset, such as, number of sensors, $n$ and edge density, $E_d$ to do this. \textit{AutoSubGraphSample} recommends $P_{haar}$ in Phase-I for any edge density in smaller networks (when $n$ is less than $90$) and high edge density (when $E_d$ is greater than $0.40$) in large networks (when $n$ is greater than $90$). It recommends $P_{nei}$ in Phase-I for large networks (when $n$ is greater than $90$) with low edge density (less than $0.40$). It recommends \textit{SMMR} or \textit{Frob} when edge density is low and \textit{SRel} or \textit{Frob}, otherwise. Our decision of the threshold for $n$ as $90$ and $E_d$ as $0.40$ is based on our observations from our experiments which we discuss in Section~\ref{s:Exp}. The pseudocode of \textit{AutoSubGraphSample} is shown in Algorithm \ref{alg:senRecc1}. 
\begin{algorithm}
\caption{AutoSubGraphSample : Recommendation for Phase-I and Phase-II}\label{alg:senRecc1}
\begin{algorithmic}[1]
\State Input: The set of sensors, $\mathcal{S}={S_1,S_2,\ldots,S_n}$ where $S_{1T}$ is the time series for sensor, $S_1$
\State Let, $e_d$ be the desired edge density
\If{{$n$ < $90$}}
%
\State  Use $P_{haar}$ in Phase-I
\If{{$e_d$ < $0.40$} }
\State {Use \textit{SMMR} or \textit{Frob} in Phase-II}
\Else
\State{Use \textit{SRel} or \textit{Frob} in Phase-II}
\EndIf
\Else
\If{{$e_d$ < $0.40$} }
\State Use $P_{nei}$ in Phase-I 
\State {Use \textit{SMMR} or \textit{Frob} in Phase-II}
\Else
\State Use $P_{haar}$ in Phase-I 
\State{Use \textit{SRel} or \textit{Frob} in Phase-II}
\EndIf
\EndIf
\end{algorithmic}
\end{algorithm}


\section{Experimental Setup}\label{s:Exp}

In this Section, we describe the datasets used in experiments and discuss the different evaluation metrics. 

\subsection{Dataset Details and Preprocessing}~\label{s:data}
The datasets used for our experiments are:
\begin{itemize}
\item \textit{$D_{epa}$:} This dataset comprises of $92$ sensors and their edge relationships which is simulated with EPANET \cite{rossman2000epanet}. EPANET is a tool for simulating water distribution network. 
\item \textit{$D_{temp}$:}  This dataset is based on a  sensor network that comprises of $74$ sensors and their hourly \textit{temperature} \cite{diamond2013us}. 

\item  \textit{$D_{pol}$:}  This dataset is based on a sensor network deployed at \textit{Aarhus, Denmark} that comprises of $37$ sensors and their \textit{Ozone level} recording \footnote{http://iot.ee.surrey.ac.uk:8080/datasets/pollution/index.html}.

\item \textit{$D_{ws}$:}  We create a \textit{synthetic dataset} of $100$ nodes that follows the Watts-Strogatz Model \cite{watts1998collective} with $\beta=0.5$. We create the data for each of the sensors such that it is strictly bandlimited in the graph Fourier domain.

\end{itemize}

\subsection{Evaluation Metrics}\label{s:metr}

\par In this Subsection, we discuss the different metrics that we used to compare the different approaches for Phase-I, Phase-II and their combinations. For Phase-I, we compare the approaches in creating different graph topology given a dataset through \textit{average path length}, \textit{clustering coefficient}, \textit{edge density} and measure how well the graph topology represents the dataset by \textit{total cumulative energy residual}. Furthermore, we use \textit{reconstruction error} to measure the performance of Phase-II and the combination of both. We do not discuss \textit{average path length}, \textit{clustering co-efficient} and \textit{edge density} further as they are well known. We detail how we calculate \textit{reconstruction error} and \textit{total cumulative energy residual} next.

\subsubsection{Reconstruction Error} ~\label{s:recon}
Reconstruction of signals on graphs is a well-known problem \cite{wang2015local,narang2013signal} that provides an estimation of the whole graph, $\mathcal{G}$ by a \textit{representative sampling subset}. For our experiments, we compare the sampling techniques on the basis of the reconstruction error. We discuss next how we calculate the reconstruction error for each sampling technique. Given the $OSP$ which comprises of $K$ \textit{representative sampling subsets} and $t$ as the length of the time series, we calculate the reconstruction error of a \textit{representative sampling subset} of $OSP$, $\mathcal{SP}_{p}$, with the $\mathcal{G}$ by measuring the difference between the signal generated by $\mathcal{SP}_{p}$, $\hat{x}$, with respect to the signal of $\mathcal{G}$, $x$, at a time-stamp, say $k$ as  
\begin{equation}
\Vert x^k-\hat{x}^k\Vert_2.
\end{equation}
We repeat this for all $K$ and $t$ respectively for each sampling technique. We calculate the reconstruction error of $OSP$, $Err(OSP,S)$ as the average of the total reconstruction error ($TErr(OSP,S)$) over $K$ \textit{representative sampling subsets}. $TErr(OSP,S)$ is the sum of the average reconstruction error of each \textit{representative sampling subset}, i.e., $\mathcal{SP}_{p}$ such that p ranges between $1$ to $K$. We calculate the reconstruction error of a \textit{representative sampling subset}, $\mathcal{SP}_{p}$ as $SamSErr(\mathcal{SP}_{p},S)$ over $t$ time stamps. Therefore, we calculate $Err(OSP,S)$ as follows :

\begin{equation}
\begin{aligned}
SamSErr(\mathcal{SP}_{p},S)=\sum_{m=1}^t\frac{(\Vert x^m-\hat{x}^m_p\Vert_2)}{\Vert x^m\Vert} \\
TErr(OSP,S)=\sum_{p=1}^K(SamSErr(\mathcal{SP}_{p},S)/t) \\
Err(OSP,S)=\frac{TErr(OSP,S)}{K}
\end{aligned}
\end{equation}

For our results, we show the quartile of $TErr(OSP,S)$ which represents the reconstruction error by a sampling technique.

\subsubsection{Total cumulative energy residual, $TCER$} ~\label{s:tcer}$TCER$ \cite{kalofolias2017learning} measures the expected energy given a data set to understand how the graph structure represents the data by total cumulative energy of the data. Total cumulative energy of the data is measured by : 
\begin{equation}
    \mathcal{T}(X,Q)=\sum_{r_=1}^N(N+1-r)\Vert q_r^\top X\Vert^2
\end{equation}
where $Q$ is an orthogonal basis, $TCER$ can then be calculated as
\begin{equation}
    1-\frac{\mathcal{T}(X,V)}{\sum_{R=1}^N\sum_{r=1}^R\Sigma_{rr}^2}
\end{equation}
where $V$ is the eigen vectors of the graph Laplacian and $\Sigma$ is the singular values of $X$. The values of $TCER$ are in the range of $[0,1]$ where a high value indicates that the dataset is well represented by the graph and a low value indicates it is not. We follow \cite{perraudin2014gspbox} for the implementation.

\begin{table*}[]
 \begin{center}
   \begin{tabular}{|l|l|l|l|l|l|l|l|l|l|l|l|}
\hline
      \textbf{Phase-I} & \textbf{{$E_d$}} & \textbf{{Avg $P_l$}} & \textbf{{Avg CC}} & \textbf{TCER} & \textbf{Th} & \textbf{Phase-I} & \textbf{{$E_d$}} & \textbf{{Avg $P_l$}} & \textbf{{Avg CC}} & \textbf{TCER} & \textbf{Th}\\
      \hline
{\multirow{ 4}{*}{$G_{dtw}$}} &  {{0.19}} & {{inf}} & {{0.80}} & 0.95 & {{20}} & {\multirow{ 4}{*}{$G_{haar}$}}  &  {{0.18}} & {{inf}} & {{0.65}} & 0.98 & {{16}}\\
       & {{0.41}} & {{inf}} & {{0.83}} & 0.96 & {{55}} &  & {{0.40}} & {{inf}} & {{0.75}} & 0.97 & {{18}}\\
    &  {{0.60}} & {{inf}} & {{0.85}} & 0.98 & {{80}} & &  {{0.60}} &  {{1.41}} & {{0.82}} & 0.96 & {{70}}\\
    & {{0.76}} & {{1.31}} & {{0.91}} & 0.97 & {{120}} &  & {{0.75}} & {{0.87}} & {{1.25}} & 0.88 & {{160}}\\
   \hline
    {\multirow{ 4}{*}{$G_{nei}$}} &  {{0.20}} & {{1.79}} & {{0.81}} & 0.77 & {{8}} & {\multirow{ 4}{*}{$G_{gsp}$}} &  {{0.39}} & {{1.61}} & {{0.87}} & 0.92 & {{$-1.111\cdot10^{-6}$}}\\
     & {{0.40}} & {{1.59}} & {{0.84}} & 0.79 & {{17}}  &  &  {{0.22}} & {{1.78}} & {{0.90}} & 0.93 & {{$-1\cdot10^{-7}$}}\\
      &  {{0.60}} & {{1.38}} & {{0.84}} & 0.78 & {{28}} &  &  {{0.59}} & {{1.41}} & {{0.89}} & 0.90 & {{$-1.1559\cdot10^{-6}$}} \\
       & {{0.75}} & {{1.247}} & {{0.87}} & 0.90 & {{37}} &  &  {{0.77}} & {{1.23}} & {{0.96}} & 0.92 & {{$-1.1576\cdot10^{-6}$}}\\
  \hline
   \end{tabular}
      \end{center}
      \caption{$E_d$, Avg $P_l$, Avg $CC$ for different values of threshold for Phase-I algorithms for \textit{$D_{temp}$} is shown}
      \label{tab:data2}
 \end{table*}

  \section{Results and Discussions}~\label{s:res}
In this Section, we initially evaluate the performance of the approaches for Phase-I and Phase-II separately followed by the validation of Algorithm \textit{AutoSubGraphSample} on $4$ representative datasets. We also analyze which combination of algorithms for Phase-I and Phase-II provides most optimal solutions. Lastly, we evaluate the performance of \textit{SubGraphSample} when the whole time series is not available.


\subsection{Phase -I Results: Comparison of the Similarity Graph Creation Approaches}~\label{s:exp01}
We evaluate the Phase-I algorithms by analyzing two specific properties of the similarity graph topology, $TCER$ and \textit{reconstruction error}.


\subsubsection{Evaluation of the Similarity Graph Topology}
~\label{s:exp8}
In order to analyze the properties of the graphs created by different graph creation approaches, we vary the values of the threshold for each of approaches of Phase-I to create graphs with a specific $E_d$ and then, study the \textit{average path length}, and \textit{clustering coefficient} of these graphs. For our experiments, we consider $4$ different edge densities; $0.20$, $0.40$, $0.60$ and $0.75$ for all the datasets. We show few representative observations for \textit{$D_{temp}$} in Table \ref{tab:data2}. Our observations show that there is a significant variance in the properties of the graphs created by the different approaches even for the same \textit{$E_d$} and same \textit{dataset}.

\par \textbf{$G_{dtw}$} is disconnected when the \textit{$E_d$} is less than $0.70$ and the number of sensors is greater than $70$ and when the number of sensors is greater than $90$ for any $E_d$. $G_{haar}$ is disconnected when the $E_d$ is less than $0.40$ and the number of sensors is above $90$ and $G_{nei}$ is always connected irrespective of the number of sensors and $E_d$. Additionally, analyzing the possible values of threshold for different edge densities, we observe  $P_{dtw}$, $P_{haar}$ and $P_{nei}$ can create graphs with any $E_d$. However, a very small difference in the values of threshold for $P_{gsp}$, $P_{corr}$ and $P_{deconv}$ can create graphs with highly different $E_d$. As previously discussed in Section \ref{s:gsp}, we observe that it is difficult for $P_{gsp}$ to generate graphs of different edge densities given a dataset.

\par The reason for the performance of $P_{corr}$ and $P_{deconv}$ is that they utilize correlation of the time series between a pair of sensors to create an edge and find similarity even when the values of the two time series vary. Therefore, we do not consider $P_{corr}$ and $P_{deconv}$ henceforth. On the basis of our observations, we find that $P_{nei}$ can be used irrespective of the dataset and $E_d$, $P_{haar}$ can be used only for datasets with small number of sensors or sensors when $E_d$ is greater than $0.40$ while $P_{dtw}$ can be used for small networks. $P_{gsp}$ can be used only if the threshold is tuned for different edge densities.

\subsubsection{Total cumulative energy residual} ~\label{s:exp8b}
We compare the $TCER$ value of a graph to that of a random graph for a dataset. Our observations indicate that $G_{dtw}$ and $G_{haar}$ always yield the best $TCER$ values, around $0.99-0.95$ irrespective of the $E_d$ and the dataset and $G_{nei}$ has the lowest $TCER$ values. We show our observations in Table \ref{tab:data2}. We observe that the $TCER$ for $D_{ws}$ is bad irrespective of the graph creation approach. The reason for this is that the $TCER$ value for the original graph from which we simulate the data for $D_{ws}$ is much lower than $1$, i.e., $0.89$.

\subsubsection{Summary for Phase-I}~\label{s:sumPhase1} 
We conclude that $P_{haar}$ followed by $P_{dtw}$ is the best choice for Phase-I for a dataset with more than $90$ nodes and high $E_d$ (more than $0.40$) and for any $E_d$ for a dataset with less than $90$. However, for graphs with more than $90$ nodes and $E_d$ less than $0.40$, we recommend $P_{nei}$ followed by $P_{gsp}$. We use these observations to propose \textit{AutoSubGraphSample}. As already discussed, we do not recommend $P_{corr}$ and $P_{deconv}$.

\begin{figure*}[ht]
     \centering
     \subfigure[]{\label{fig:res2a}\includegraphics[width=2in]{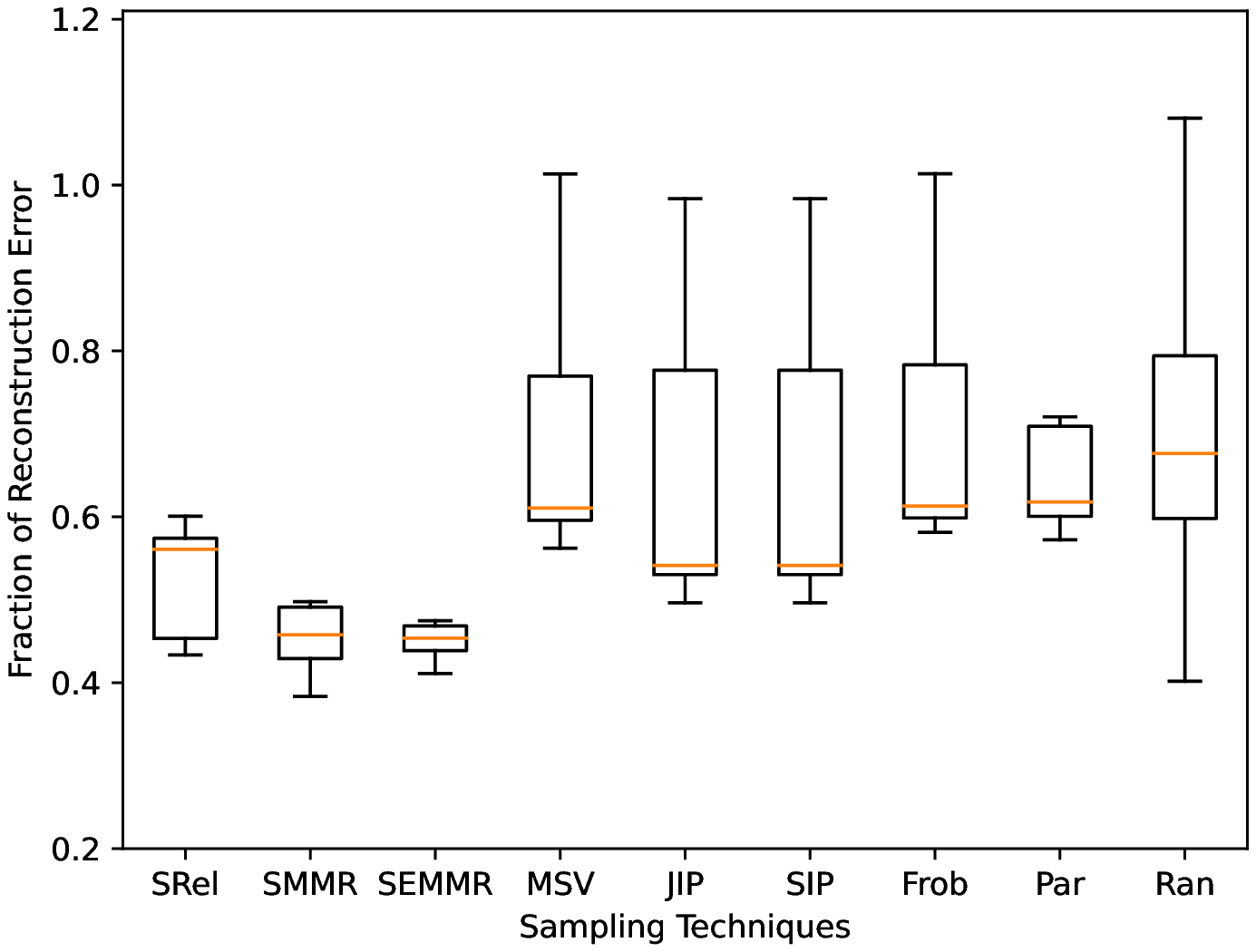}}
    \subfigure[]{\label{fig:res2b}\includegraphics[width=2in]{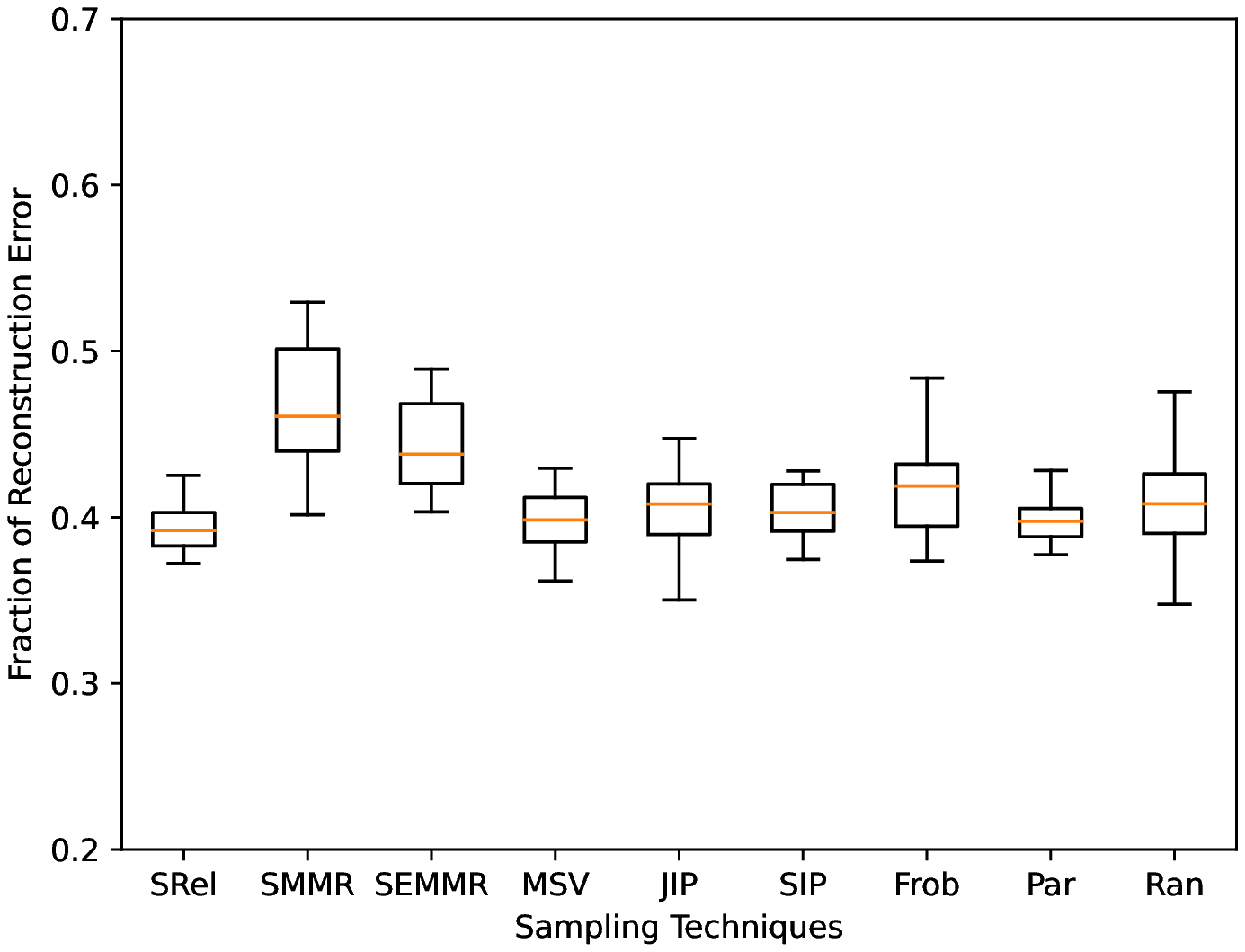}}
  \subfigure[]{\label{fig:res2c}\includegraphics[width=2in]{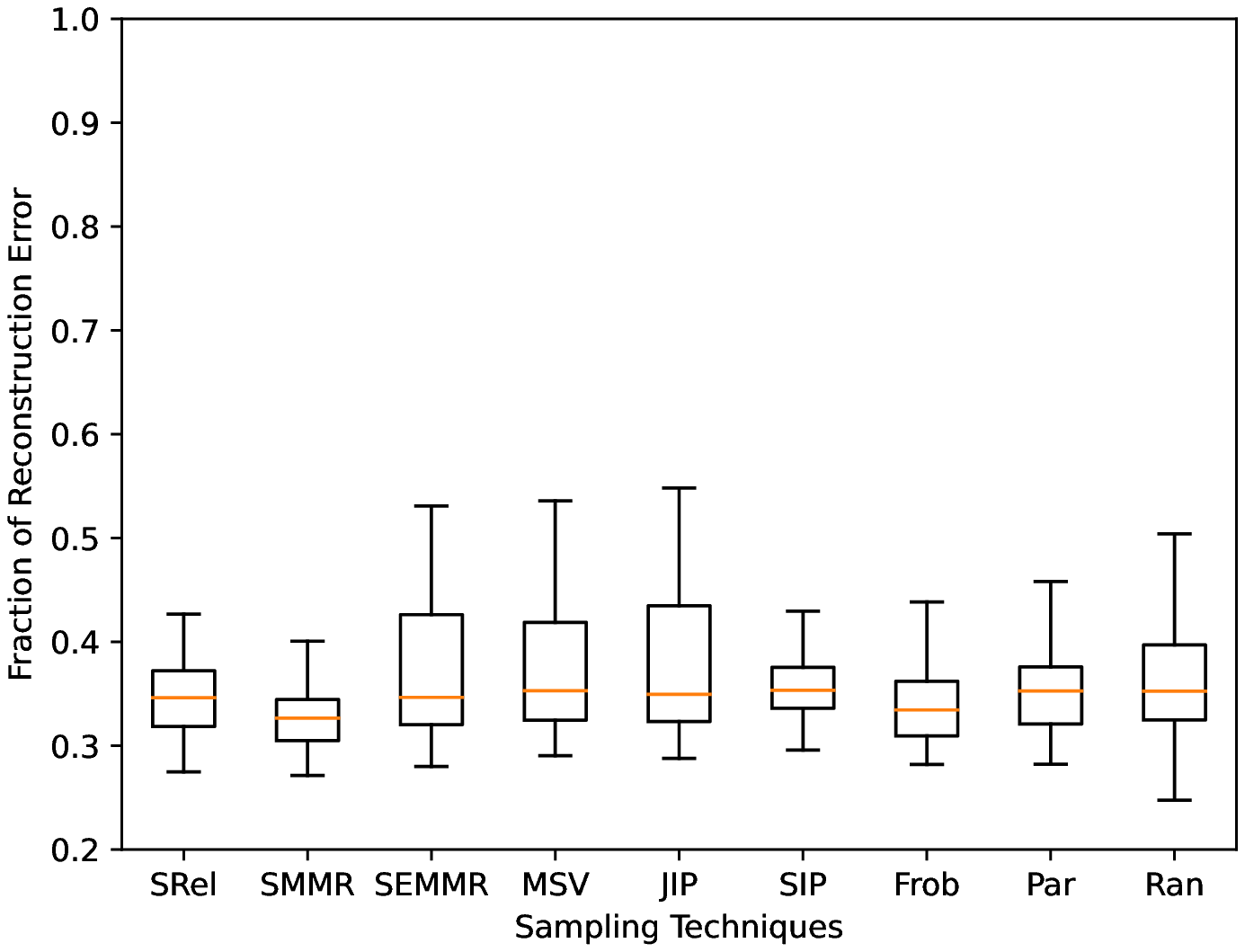}}
      \caption{Comparing the reconstruction error of sampling techniques and Random Sampling Approach for $K$ is $5$, $E_d$ is $0.20$ on $G_{nei}$ in Figure \ref{fig:res2a} of \textit{$D_{temp}$}, for $K$ is $7$ when $E_d$ is $0.75$ on $G_{haar}$ in Figure \ref{fig:res2b} of \textit{$D_{temp}$} and $K$ is $5$, $E_d$ is $0.75$ on $G_{haar}$ in Figure \ref{fig:res2c} of \textit{$D_{pol}$}. The x-axis represents the sampling techniques and y-axis represents the reconstruction error.}
      \label{fig:phase2Res1}
\end{figure*}

\subsection{Phase-II Results: Comparison of the Sampling Techniques}~\label{s:exp1}
We now compare the performance of the sampling approaches for Phase-II and a random \textit{representative sampling subset} selection algorithm on the basis of their solution for Equation \ref{eq:probForm1}. Therefore, we compare the reconstruction error generated by the sampling techniques for each graph creation approach with different $E_d$ as $0.20$, $0.40$, $0.60$ and $0.75$ and vary $K$ from $5-13$ and calculate the \textit{reconstruction error} quartile. 
\par We find that irrespective of $K$ and the dataset, \textit{SRel} ranks $1-3$ among all sampling techniques when the $E_d$ is greater than $0.40$ whereas \textit{SMMR} ranks $1-3$ when the $E_d$ is less than $0.40$. \textit{SEMMR} has similar mean \textit{reconstruction error} as \textit{SMMR} but, the maximum \textit{reconstruction error} is much higher. \textit{SRel} has around $0.1-0.40$ reconstruction error when $E_d$ is greater than $0.40$ and $0.20-0.60$ otherwise. \textit{SMMR} and \textit{SEMMR} has around $0.2-0.40$ when $E_d$ is less than $0.40$ and $0.10-0.60$ otherwise. \textit{SRel}, \textit{SMMR} and \textit{SEMMR} has the highest maximum reconstruction error for $D_{epa}$ when $E_d$ is greater than $0.60$ and is $K$ greater than $7$.
\textit{MSV} has around $0.20-0.40$ reconstruction error when $E_d$ is around $0.75$ but the performance degrades for low $E_d$ to around $0.30-0.80$ reconstruction error. \textit{MSV} also has the highest maximum reconstruction error at low $E_d$. On comparing \textit{JIP} and \textit{SIP} which follow similar approaches, we observe that \textit{SIP} yields better performance than \textit{JIP} in every scenario irrespective of $K$, $E_d$ or dataset. On comparison with the other sampling approaches, we observe that \textit{SIP} has around $0.40-0.60$ reconstruction error when $E_d$ is high and $0.40-0.90$ otherwise. \textit{Par} has the worst performance among all sampling techniques. Although \textit{Frob} ranks in the top $3-4$ among all sampling techniques based on the minimum \textit{reconstruction error}, it produces the maximum reconstruction error among all sampling techniques. As expected, we also observe that the reconstruction error increases with increase in $K$ irrespective of the sampling technique. 

\par Based on our observations, we conclude that \textit{SRel} is the best choice for graphs with high $E_d$ (greater than $0.40$) and \textit{SMMR} for graphs with $E_d$ less than $0.40$. However, if we need to choose sampling technique that performs irrespective of the $E_d$, \textit{Frob} should be selected. We use these observations to propose \textit{AutoSubGraphSample}. Due to the huge number of results, we only show $3$ representative examples in Figure \ref{fig:phase2Res1}.

\begin{table*}[]
 \begin{center}
   \begin{tabular}{|l|l|l|l|l|l|l|l|l|l|l|l|}
\hline
      \textbf{Dataset} & \textbf{{$E_d$}} & \textbf{\textit{AutoSubGraphSample}} & \textbf{\textit{Manual Selection}} & \textbf{Dataset} & \textbf{{$E_d$}} & \textbf{\textit{AutoSubGraphSample}} & \textbf{\textit{Manual Selection}}\\
      \hline
{\multirow{ 2}{*}{$D_{ps}$}} &  {{0.25}} & {{0.51}} & {{0.51}} & {\multirow{ 2}{*}{$D_{gas}$}}  &  {{0.25}} & {{0.49}} & {{0.47}} \\
       & {{0.75}} & {{0.61}} & {{0.61}} &  & {{0.75}} & {{0.56}} & {{0.53}}\\
    
   \hline
    {\multirow{ 2}{*}{$D_{hum}$}} &  {{0.25}} & {{0.31}} & {{0.25}} & {\multirow{ 2}{*}{$D_{in}$}}  &  {{0.25}} & {{0.33}} & {{0.32}} \\
       & {{0.75}} & {{0.45}} & {{0.42}} & & {{0.75}} & {{0.49}} & {{0.47}}\\
  \hline
   \end{tabular}
      \end{center}
      \caption{Comparing the reconstruction error by \textit{AutoSubGraphSample} and \textit{Manual Selection} on $D_{ps}$, $D_{gas}$, $D_{in}$ and $D_{hum}$ when $E_d$ is $0.25$ or $0.75$ for K = $5$}
      \label{tab:manAut0}
 \end{table*}

\subsection{Evaluation of AutoSubGraphSample}~\label{s:exp15R} 
Based on our observations for Phase-I and Phase-II, we decide the values for $Th_n$ and $Th_e$ in Algorithm \textit{AutoSubGraphSample} as $90$ and $0.40$ respectively. We analyze the generalizability of \textit{AutoSubGraphSample} on $4$ new representative datasets now. 

\begin{itemize}
    \item $D_{ps}$: A dataset that records temperature of $55$ sensors.\footnote{https://archive.ics.uci.edu/ml/datasets.php}
    \item $D_{in}$:  A dataset that records humidity of $54$ sensors.\footnote{https://www.kaggle.com/hmavrodiev/sofia-air-quality-dataset?select=2017-09_bme280sof.csv}
    \item $D_{hum}$: A dataset that records humidity of $100$ sensors.\footnote{https://www.kaggle.com/hmavrodiev/sofia-air-quality-dataset?select=2017-09_bme280sof.csv}
     \item $D_{gas}$: A dataset that records acetone of $16$ sensors.\footnote{https://archive.ics.uci.edu/ml/datasets/Gas+sensor+array+under+flow+modulation}
\end{itemize}

\begin{figure*}[ht]
\centering
    \subfigure[]{\label{fig:eval4}\includegraphics[width=1.5in,angle =90 ]{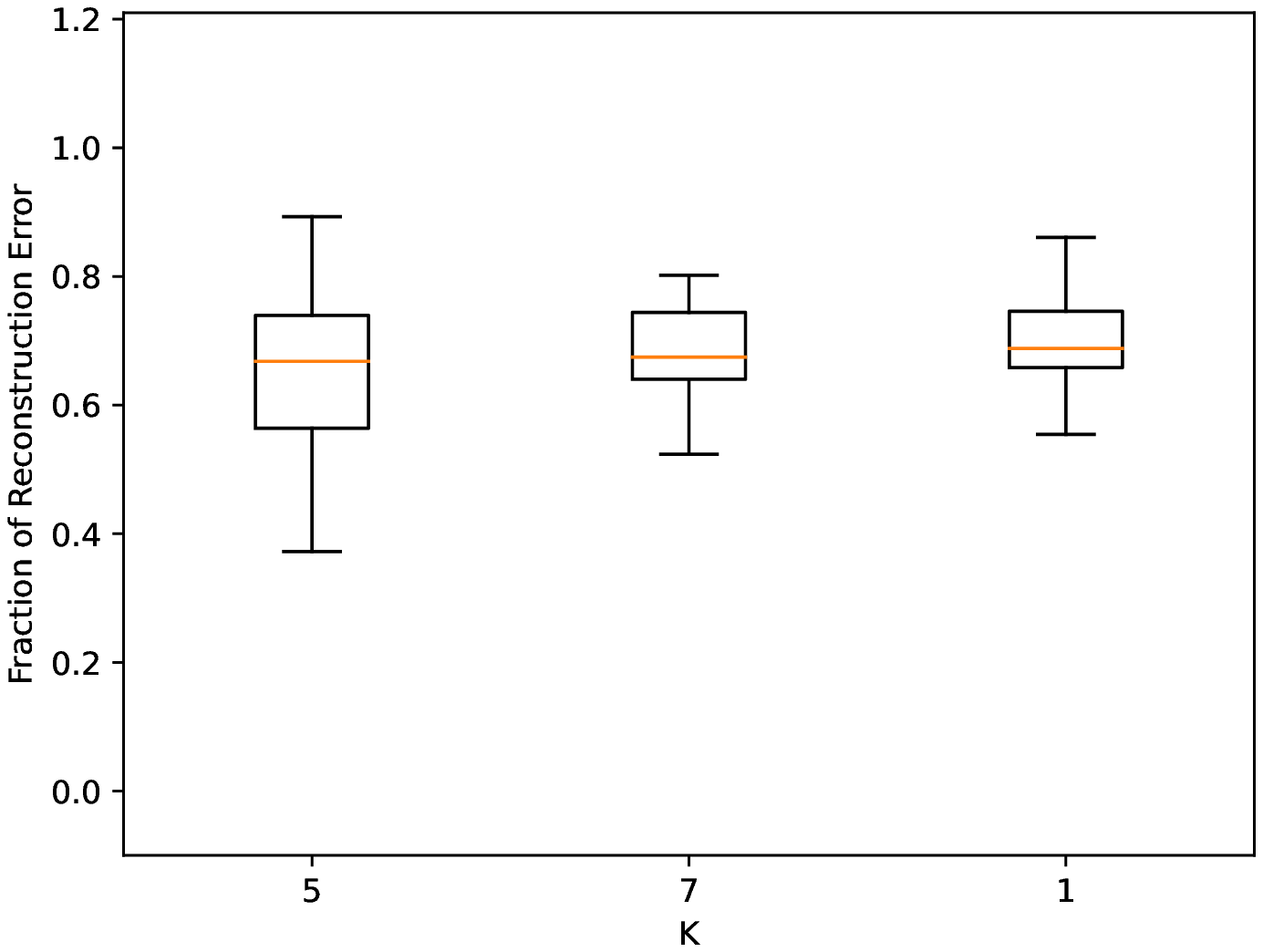}}
\subfigure[]{\label{fig:evalN4}\includegraphics[width=1.5in,angle =90 ]{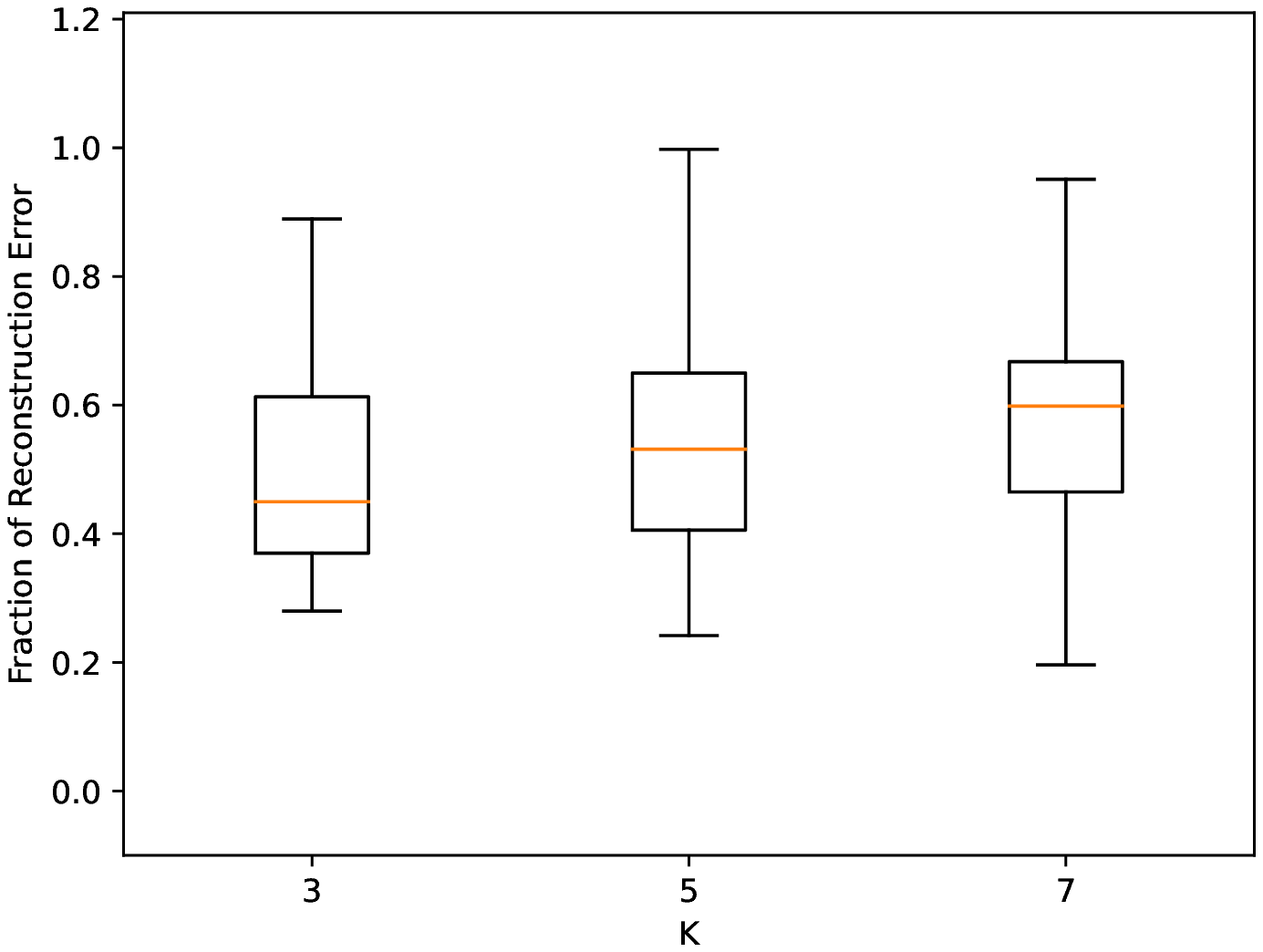}}
\subfigure[]{\label{fig:evalN2}\includegraphics[width=2in]{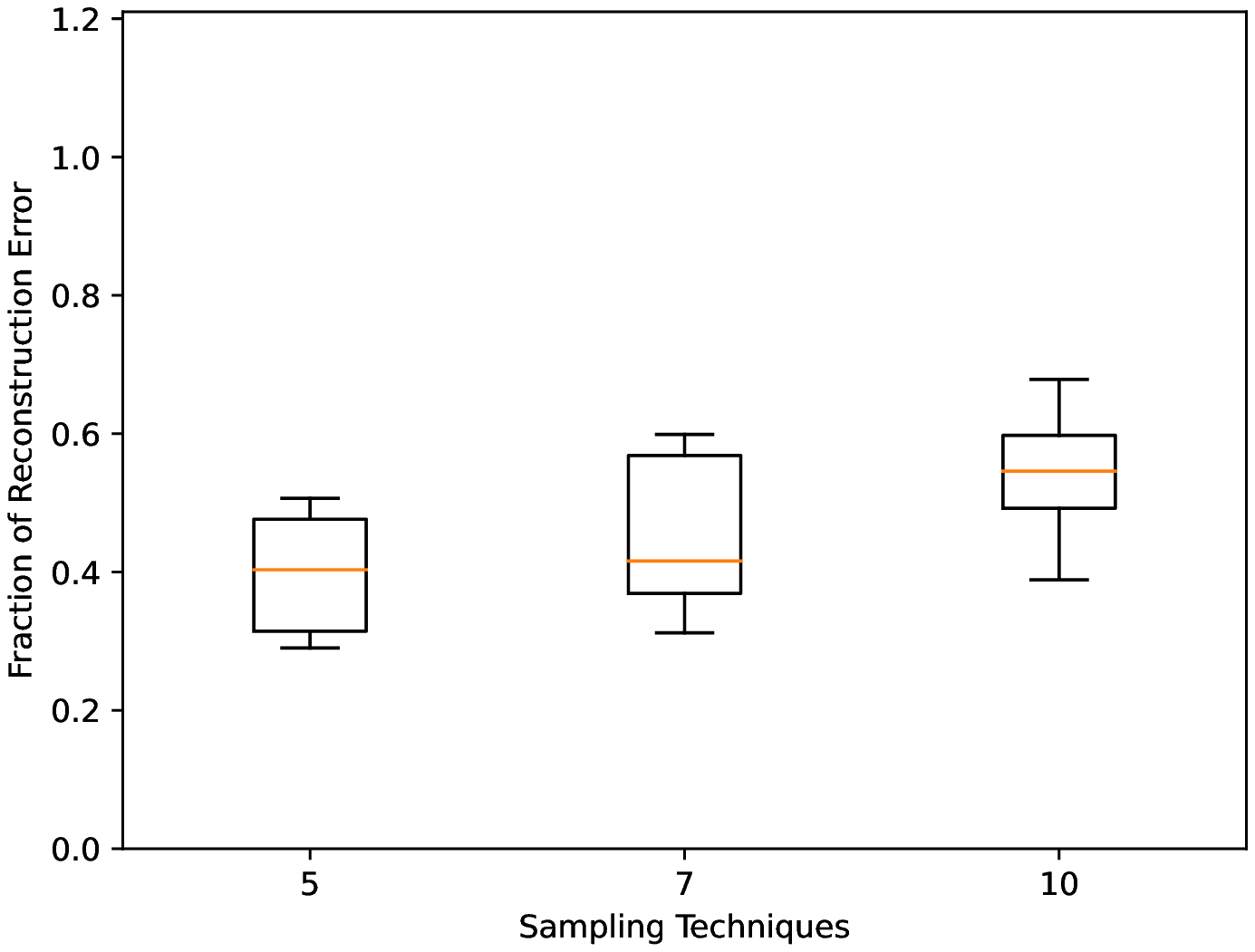}}
\caption{Comparing the reconstruction error of $D_{hum}$ when K = $5$, $7$ and $10$ in Figure \ref{fig:eval4}, $D_{gas}$ when K = $3$, $5$ and $7$ in Figure \ref{fig:evalN4} and $D_{hum}$ when K = $5$, $7$, $10$ in Figure \ref{fig:evalN2} for $E_d$ around $0.25$ The x-axis represents K and y-axis represents the reconstruction error.}
\label{fig:evalN1}
\end{figure*}

Based on \textit{AutoSubGraphSample}, we apply $P_{haar}$ in Phase-I irrespective of the $E_d$ and \textit{SMMR} or \textit{SRel} in Phase-II on the basis of $E_d$ for $D_{ps}$, $D_{in}$ and $D_{gas}$. However, for $D_{hum}$, we apply $P_{haar}$ in Phase-I and \textit{SRel} in Phase-II when $E_d$ is within $0-0.40$ and $P_{nei}$ in Phase-I followed by \textit{SMMR} in Phase-II otherwise. We consider the number of \textit{representative sampling subsets}, $K$ as $5$, $7$ and $10$ for $D_{ps}$, $D_{hum}$ and $D_{in}$. As $D_{gas}$ comprises of only $16$ sensors, we consider $K$ as $3$, $5$ and $7$. We observe that the reconstruction error for $D_{gas}$, $D_{hum}$, $D_{ps}$ and $D_{in}$ by \textit{AutoSubGraphSample} irrespective of the number of \textit{representative sampling subsets} and $E_d$ is similar to our previous observations for other datasets. We show the reconstruction error of $D_{ps}$, $D_{gas}$ and $D_{hum}$ in Figure \ref{fig:evalN1} when $E_d$ is $0.25$. In order to understand the significance of \textit{AutoSubGraphSample}, we compare the performance by \textit{AutoSubGraphSample} and \textit{Manual Selection}, i.e., if we manually select the best combination of Phase-I and Phase-II algorithms specifically for a dataset. We apply all combinations of Phase-I and Phase-II algorithms on a dataset and calculate the respective reconstruction errors for a specific $E_d$. We select that combination of Phase-I and Phase-II algorithm which provides the least reconstruction error as the \textit{Manual Selection}. We repeat this for $D_{ps}$, $D_{in}$, $D_{hum}$ and $D_{gas}$ when $E_d$ is $0.25$ and $0.75$ respectively for $K$ as $3$-$13$. Our observations as shown in Table \ref{tab:manAut0} for K = $5$ shows that \textit{AutoSubGraphSample} can ensure similar results as compared to \textit{Manual Selection} with a small margin of around $2-6\%$ for $D_{in}$, $D_{hum}$ and $D_{gas}$ and same results for $D_{ps}$. Therefore, based on our observations, we can conclude that \textit{AutoSubGraphSample} generalizes to a dataset irrespective of the size of the dataset and $E_d$.

\begin{figure}[ht]
     \centering
    
    \includegraphics[width=2.8in]{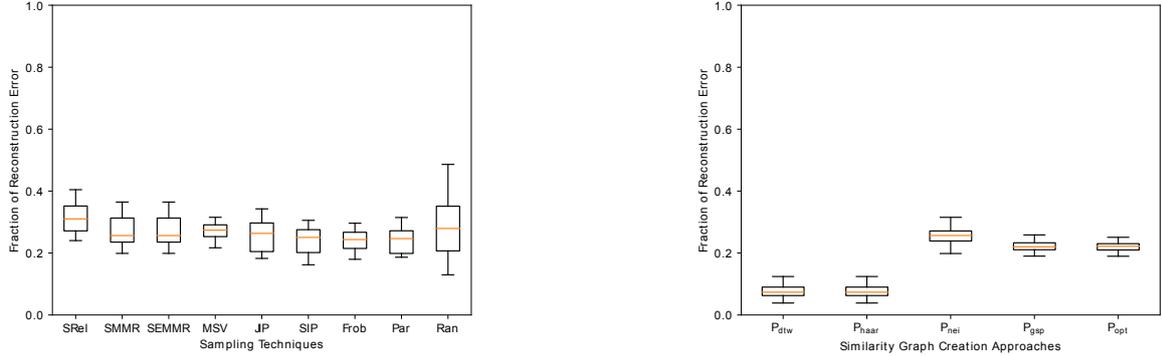}
       \label{fig:ex1}
        \caption{Comparing the reconstruction error of the sampling techniques on $G_{opt}$ for $D_{ex}$ is shown. The x-axis represents the sampling techniques and y-axis represents the reconstruction error.}
    \label{fig:ExhaustiveSearchsamplingCompare}
\end{figure}


\subsection{Comparing SubGraphSample with Exhaustive Search } \label{s:caseex}

\par In theory, identifying the optimum sampling partition, $O_s$, given the data of the sensors is possible through a joint exhaustive search for both the best graph topology and best sampling partition. However, this requires us to perform an exhaustive search for best sampling partition for every possible graph topology which is so computationally expensive that we consider it to be infeasible. Therefore, we consider this in $2$ phases, where in Phase-I, we search for an \textit{optimal graph topology} and in Phase-II, we search for the \textit{optimal sampling partition} given the \textit{optimal graph topology}. For our experiments, we consider a subset of $8$ sensors of $D_{pol}$, namely $D_{ex}$, as exhaustive analysis is not possible on the complete dataset.  
\begin{table}[]
    \centering
    \begin{tabular}{|l|l|l|l|l|}
    \hline
        \textbf{Method} & \textbf{E_d} & \textbf{{Avg $P_l$}} & \textbf{{Avg $CC$}} & \textbf{TCER}  \\
        \hline
         {{$G_{opt}$}} &  {{0.64}} & {{1.36}} & {{0.45}} & {{0.88}} \\
         {{$G_{nei}$}} &  {{0.46}} & {{1.54}} & {{0.82}} & {{0.82}}\\
         {{$G_{haar}$}} &  {{0.50}} & {{1.79}} & {{0.87}} & {{0.86}}\\
         {{$G_{dtw}$}} &  {{0.50}} & {{1.79}} & {{0.87}} & {{0.86}}\\
         {{$G_{gsp}$}} &  {{0.57}} & {{1.54}} & {{0.80}} & {{0.83}}\\
         \hline
    \end{tabular}
    \caption{E_d, $P_l$, $CC$ and TCER for $D_{ex}$ is shown}
    \label{tab:ExhaustTCER}
\end{table}

\par  In order to identify the \textit{optimal graph}, we explore the relationship between existing graph topology measures, like \textit{average path length}, \textit{clustering coefficient} and \textit{TCER} with \textit{optimal graph}. Based on our observations, we conclude that the \textit{TCER} values are indirectly proportional to \textit{reconstruction error}, i.e., higher the \textit{TCER} values, lower is the \textit{reconstruction error}. Furthermore,
if different graphs have similar \textit{TCER} values, we observe that as the \textit{average path length} decreases, the \textit{reconstruction error} also decreases. We calculate the \textit{TCER} for all possible connected graphs to a precision of $2$ significant digits for $D_{ex}$. We consider the graph which has the highest \textit{TCER} and shortest \textit{average path length} as the \textit{optimal graph}, $G_{opt}$. As the $E_d$ of $G_{opt}$ is $0.64$, we try to find graphs with similar $E_d$ using the proposed methods.
We show the $E_d$, average path length, clustering co-efficient, \textit{TCER} of $G_{opt}$ with $G_{nei}$, $G_{haar}$, $G_{dtw}$ and $G_{gsp}$ in Table \ref{tab:ExhaustTCER} which indicates that $G_{haar}$ and  $G_{dtw}$ has the most similar values \textit{TCER} with $G_{opt}$. As the graphs produced with $P_{haar}$ and $P_{dtw}$ are identical, so we only show results for $P_{dtw}$ henceforth.
\begin{figure}[ht]
    \centering
    \includegraphics[width=2.8in]{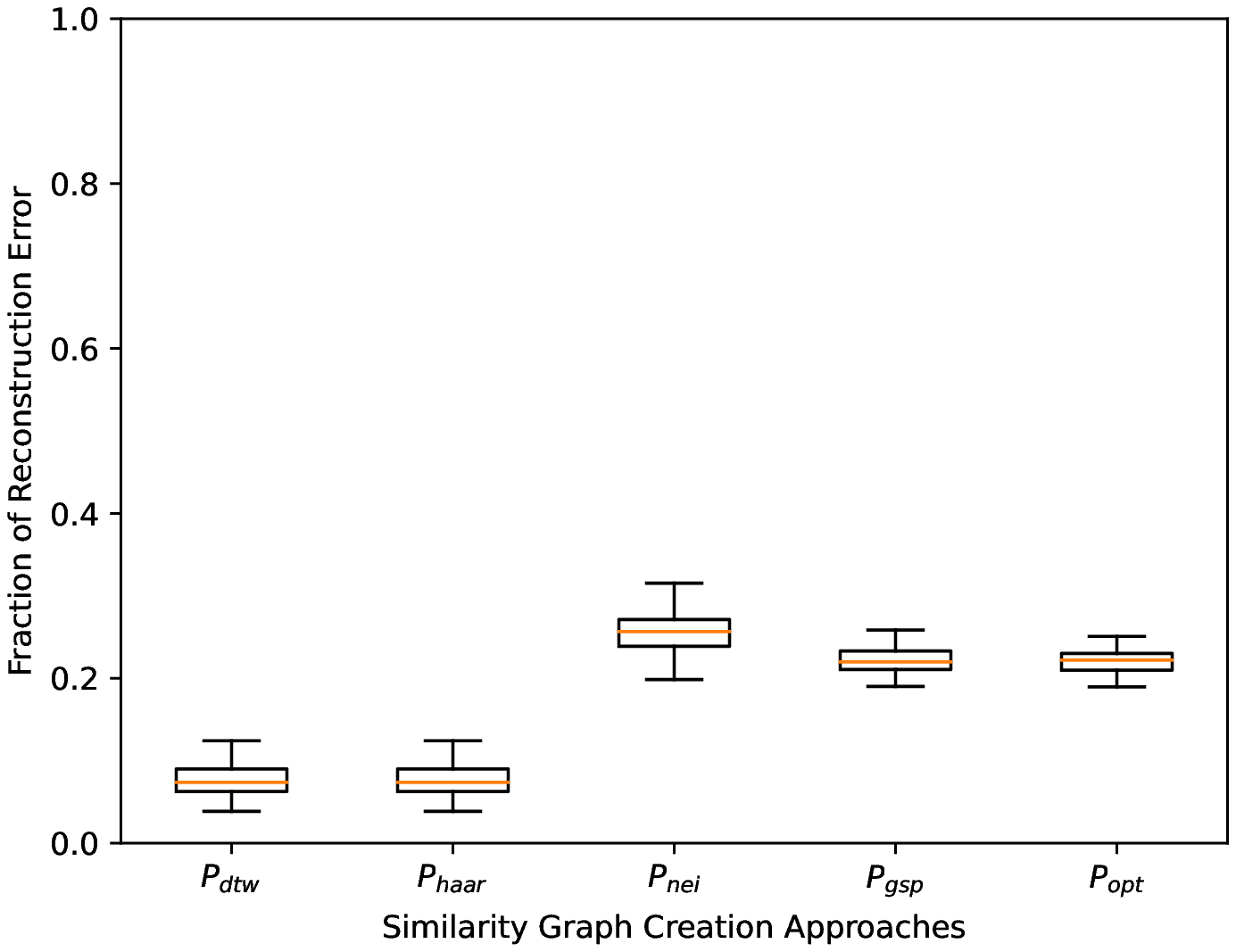}
    \caption{Comparing the reconstruction error for the optimal sampling algorithms on $G_{nei}$, $G_{haar}$, $G_{dtw}$, $G_{gsp}$, and $G_{opt}$ on $D_{ex}$ is shown. The x-axis represents the similarity graph creation approaches and y-axis represents the reconstruction error.}
    \label{fig:ExhaustiveSearchGraphCompare}
\end{figure}

\par  In order to find the $O_s$, we search all possible sampling partitions on $G_{opt}$ such that the maximum \textit{reconstruction error} is the lowest. We perform an exhaustive search to find $O_s$ on $G_{nei}$, $G_{dtw}$, $G_{gsp}$ and $G_{opt}$. Our observations as shown in Figure \ref{fig:ExhaustiveSearchGraphCompare} indicate that $P_{haar}$ and $P_{dtw}$ ensures the least reconstruction error. Therefore, our observations indicate that it is possible to find a graph that gives lower reconstruction error than the graph with the highest \textit{TCER}. To evaluate the different sampling algorithms for Phase-II, we compare the reconstruction error by \textit{Frob}, \textit{MSV}, \textit{SIP}, \textit{SMMR} and \textit{SRel} on $G_{opt}$ in Figure \ref{fig:ExhaustiveSearchsamplingCompare}. Our observations indicate that by \textit{Frob}, \textit{SIP} and \textit{SMMR} has the least reconstruction error with respect to \textit{Opt}. However, these observations varies with network size and edge density. As it is not possible to confirm every scenario of different edge densities and for different network sizes with exhaustive analysis, we compare the performance of the graph creation approaches and sampling algorithms on a synthetic dataset whose \textit{representative sampling subsets} are already provided next in  Subsection~\ref{s:cases1}.

\begin{figure}[ht]
   \includegraphics[width=2.8in]{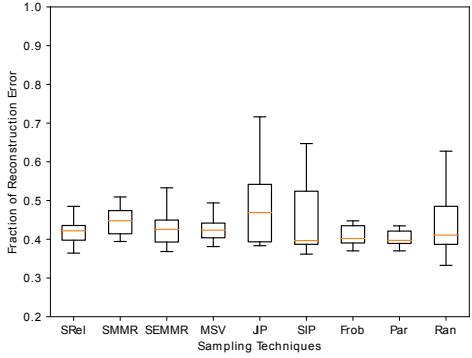}
\caption{Comparison of the reconstruction error when $E_d$ is $0.60$ and $P_{haar}$ in Phase-I for $D_{st}$ is shown.}
 \label{fig:st1}
\end{figure}

\subsection{Comparison of SubGraphSample with \textit{Optimal Sampling Sets}} ~\label{s:cases1}

We now evaluate how close the \textit{representative sampling subsets} found by \textit{SubGraphSample} are to the optimal sampling sets, $O_s$. As we do not have $O_s$ for any real dataset, we construct a dataset such that we know $O_s$. We assume the optimal number of sampling sets, $K$ as $6$, the total number of sensors, $N$ as $40$, the length of the time series as $10$, sensors as ${S_1,S_2,\ldots,S_N}$ and we denote this dataset as $D_{st}$. We simulate $D_{st}$ such that it records temperature. We generate $O_s$ of $6$, ${O_1,O_2,\ldots,O_6}$ sampling sets by randomly allocating each sensor to a $O_i$ on the basis of $D_{st}$. Based on $O_s$, we generate the time series of $S$ such that while the mean values of the distributions vary by $3-5$ between different sampling sets, i.e., the constructed sampling sets are indeed the optimal. We calculate the reconstruction error of $O_s$ for $D_{st}$ to understand the performance of $O_s$. As we do not know the true $E_d$ and the graph topology of $D_{st}$ which is required to calculate the reconstruction error, we consider $4$ different $E_d$, such as, $0.20$, $0.40$, $0.60$ and $0.75$ and the $4$ similarity graph creation algorithms, $P_{dtw}$, $P_{haar}$, $P_{nei}$ and $P_{gsp}$. We compare $P_{dtw}$, $P_{nei}$, $P_{gsp}$ and $P_{haar}$ on the basis graph topology, $TCER$ and the reconstruction error for all $E_d$ in Section~\ref{s:app} (Table \ref{tab:casestudy}). Our observations shows that $O_s$ has minimum reconstruction error when $E_d$ is $0.60$ and similarity graph creation approach is $P_{haar}$. On comparing the sampling techniques on $G_{haar}$ when $E_d$ is $0.60$, our observations as shown in Figure~\ref{fig:st1} indicate that \textit{SRel} produces similar reconstruction error to $O_s$. Therefore, the combination of $P_{haar}$ and \textit{SRel} can ensure most similar results to $O_s$. On the basis of our observations from Subsection \ref{s:cases1} and this Subsection, we find that the proposed recommendations for Algorithm \textit{AutoSubGraphSample} can ensure most similar results to $O_s$. For example, we observe that $P_{haar}$ in Phase-I, \textit{SMMR} or \textit{Frob} in Phase-II has the best performance. Although it is not possible to confirm every scenario by exhaustive analysis, our results from empirical analysis supports the recommendations by Algorithm \textit{AutoSubGraphSample} when $P_{nei}$ is used in Phase-I and when \textit{SRel} could be used in Phase-II.

\begin{figure*} [ht]
     \centering
    \subfigure[]{\label{fig:edge_D1}\includegraphics[width=2in]{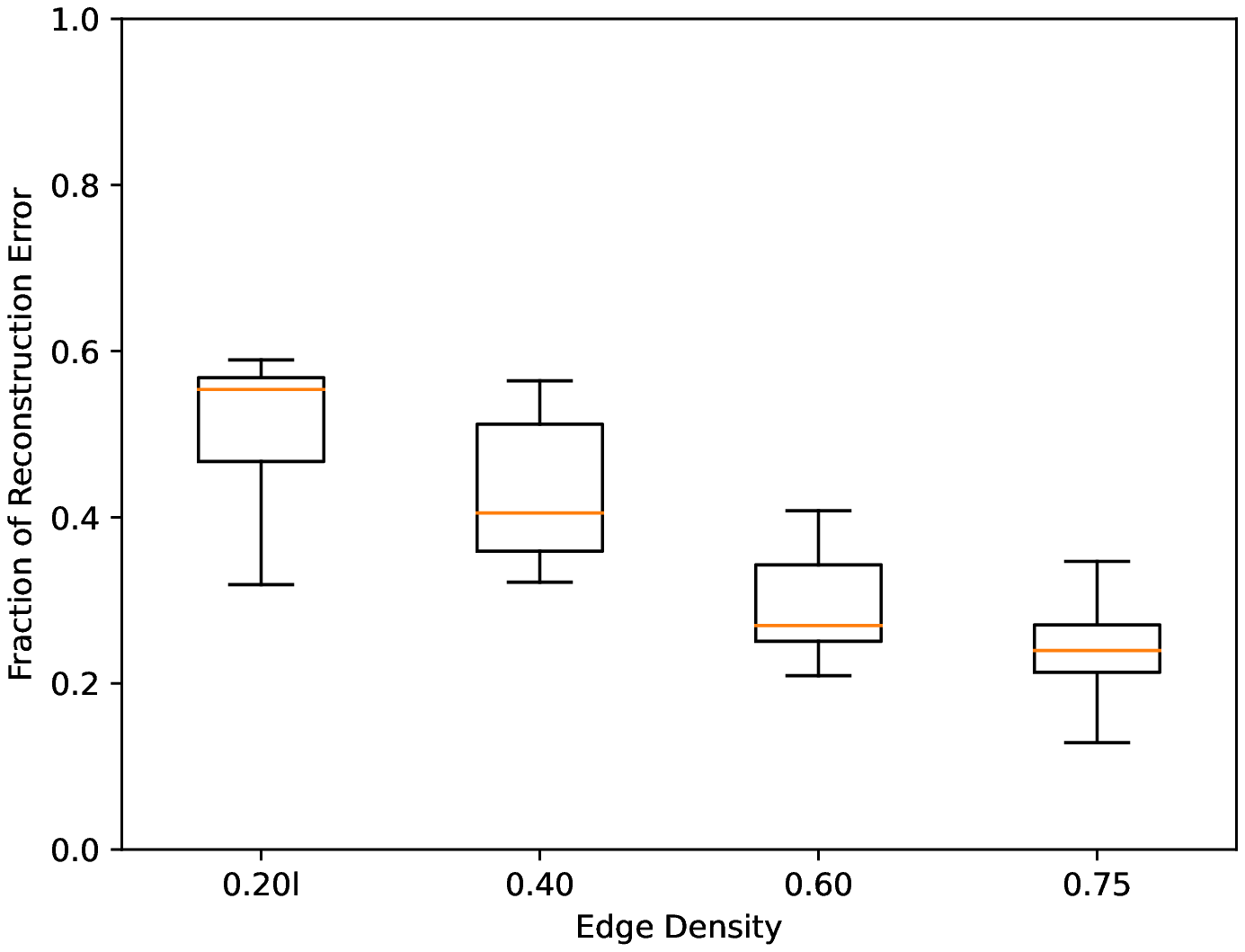}}
    \subfigure[]{\label{fig:edge_D2}\includegraphics[width=2in]{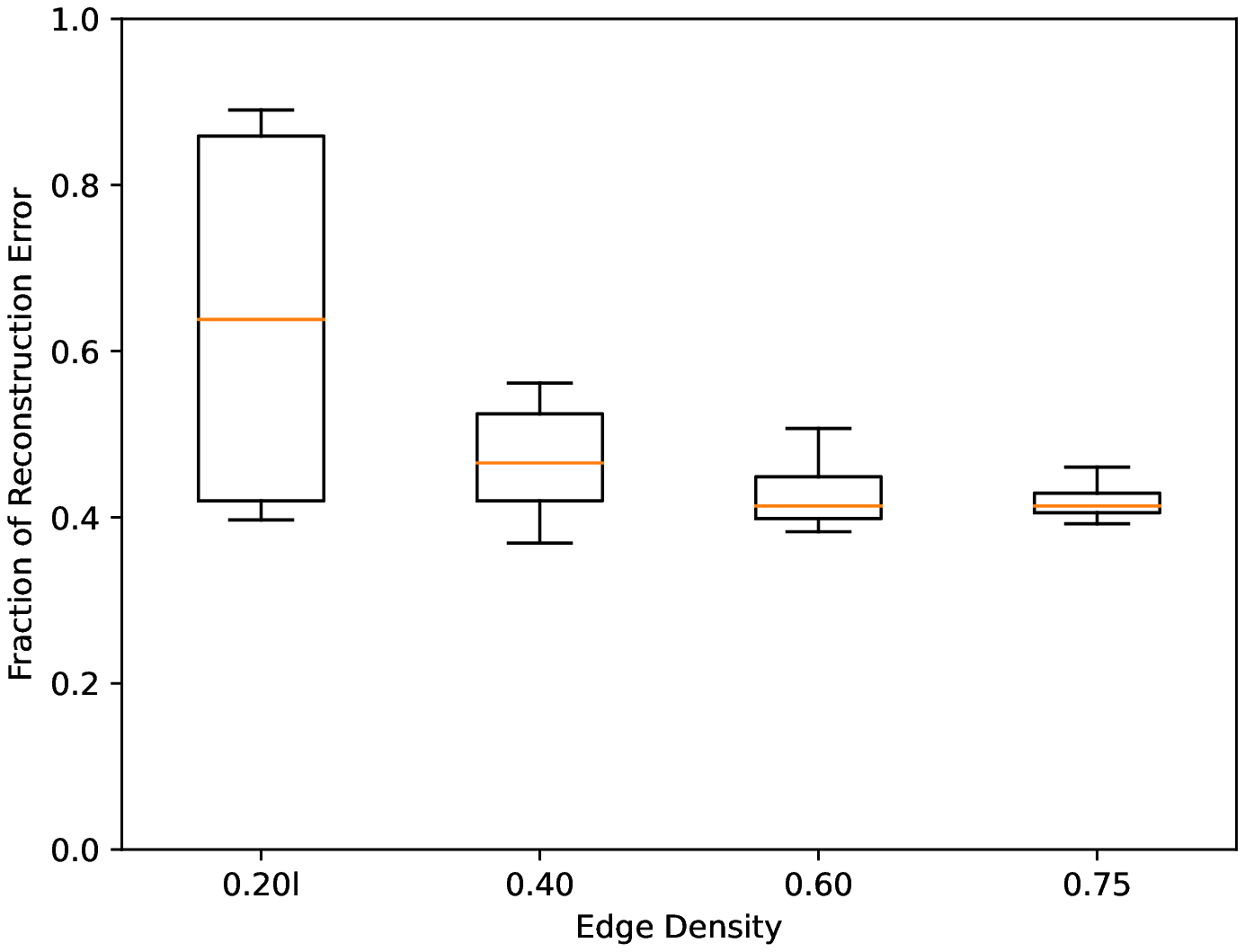}}
  \subfigure[]{\label{fig:edge_D3}\includegraphics[width=2in]{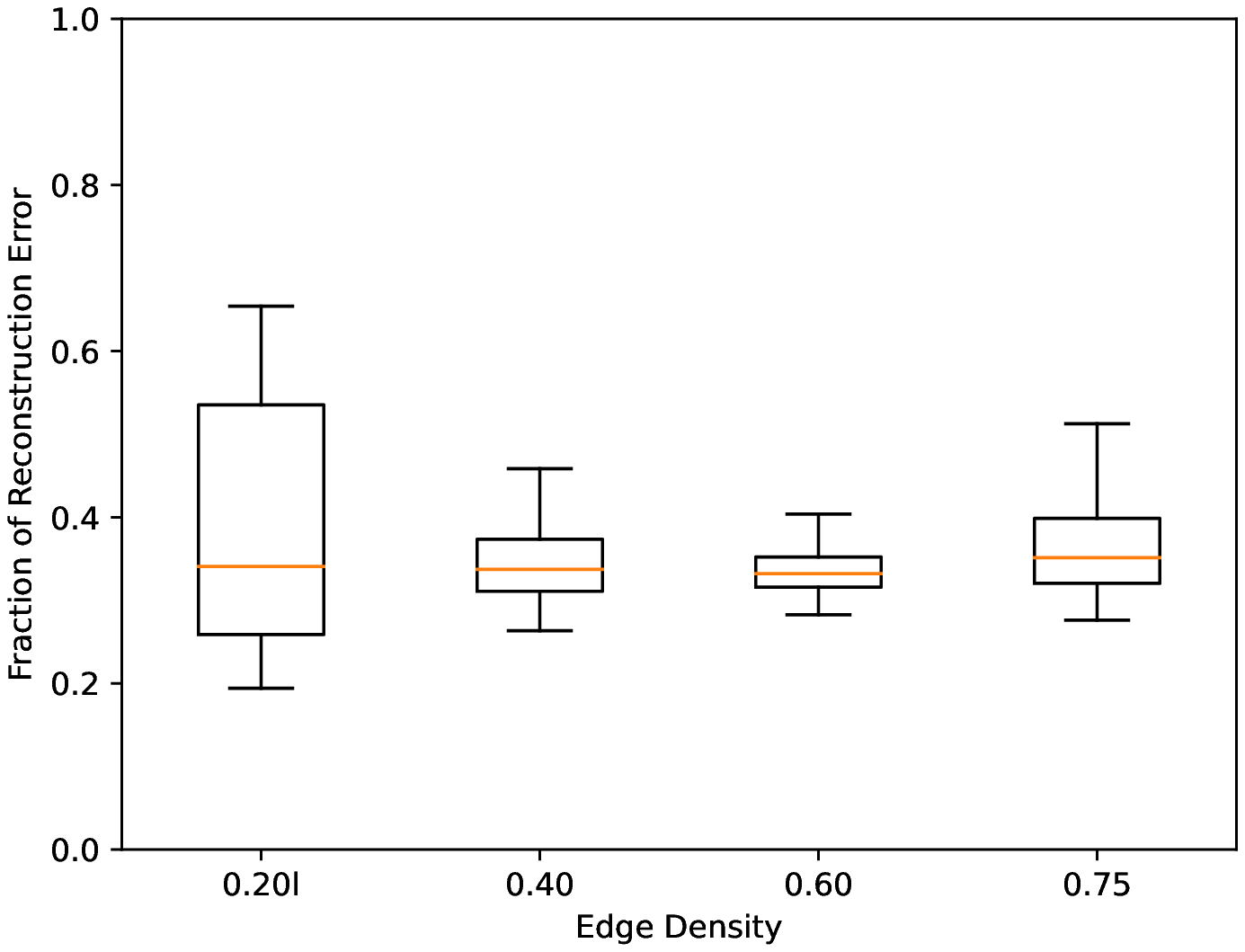}}
     
     \caption{Comparison of the reconstruction error for different edge densities for $G_{gsp}$ on $D_{epa}$ in Figure 9a, $G_{nei}$ on $D_{temp}$ in Figure 9b and $G_{dtw}$ on $D_{pol}$ in Figure 9c is shown. The x-axis represents the sampling techniques and y-axis represents the reconstruction error.}
     \label{fig:edge_density}
\end{figure*}

\begin{table*}
\begin{center}
\begin{tabular} {|l|l|l|l|l|l|l|l|l|l|l|l|}
\hline
\textbf{{Dataset}} & $\epsilon$ & \textbf{{$E_d$}} & \textbf{{Phase-I}} & \textbf{{Phase-II}} & $K$ & \textbf{{Dataset}} & $\epsilon$ & \textbf{{$E_d$}} & \textbf{{Phase-I}} & \textbf{{Phase-II}} & $K$\\
\hline
&  &  &  & \textit{SRel} & $17$ &  &  &  & & \textit{SRel} & $13$ \\
\cline{5-5} \cline{6-6} \cline{11-11}  \cline{12-12} 
$D_{temp}$ &  $250$  & $0.75$ & $G_{nei}$ & \textit{SMMR} & $15$ & $D_{temp}$ &  $250$  & $0.75$ & $G_{haar}$ & \textit{SMMR} & $11$ \\
\cline{5-5}  \cline{6-6} \cline{11-11}  \cline{12-12} 
&  &  &  & \textit{SEMMR} & $14$ &  &  &  & & \textit{SEMMR} & $11$ \\
\cline{5-5}  \cline{6-6} \cline{11-11}  \cline{12-12} 
\hline&  &  &  & \textit{SRel} & $12$ &  &  &  & & \textit{SRel} & $12$ \\
\cline{5-5} \cline{6-6} \cline{11-11}  \cline{12-12} 
$D_{temp}$ &  $250$  & $0.20$ & $G_{nei}$ & \textit{SMMR} & $11$ & $D_{temp}$ &  $250$  & $0.40$ & $G_{haar}$ & \textit{SMMR} & $13$ \\
\cline{5-5}  \cline{6-6} \cline{11-11}  \cline{12-12} 
&  &  &  & \textit{SEMMR} & $11$ &  &  &  & & \textit{SEMMR} & $13$ \\
\cline{5-5}  \cline{6-6} \cline{11-11}  \cline{12-12} 
\hline
&  &  &  & \textit{SRel} & $15$ &  &  &  & & \textit{SRel} & $14$ \\
\cline{5-5} \cline{6-6} \cline{11-11}  \cline{12-12} 
$D_{pol}$ &  $250$  & $0.75$ & $G_{nei}$ & \textit{SMMR} & $15$ & $D_{pol}$ &  $250$  & $0.75$ & $G_{dtw}$ & \textit{SMMR} & $13$ \\
\cline{5-5}  \cline{6-6} \cline{11-11}  \cline{12-12} 
&  &  &  & \textit{SEMMR} & $14$ &  &  &  & & \textit{SEMMR} & $13$ \\
\cline{5-5}  \cline{6-6} \cline{11-11}  \cline{12-12} 
\hline
&  &  &  & \textit{SRel} & $11$ &  &  &  & & \textit{SRel} & $11$ \\
\cline{5-5} \cline{6-6} \cline{11-11}  \cline{12-12} 
$D_{pol}$ &  $250$  & $0.20$ & $G_{nei}$ & \textit{SMMR} & $11$ & $D_{pol}$ &  $180$  & $0.20$ & $G_{dtw}$ & \textit{SMMR} & $12$ \\
\cline{5-5}  \cline{6-6} \cline{11-11}  \cline{12-12} 
&  &  &  & \textit{SEMMR} & $11$ &  &  &  & & \textit{SEMMR} & $13$ \\
\cline{5-5}  \cline{6-6} \cline{11-11}  \cline{12-12} 
\hline
\end{tabular}
\end{center}
\renewcommand\thetable{6}
\caption{We show the number of sampling sets, $K$ generated by \textit{SRel}, \textit{SMMR} and \textit{SEMMR} when the mean reconstruction error, $\epsilon$, is given for \textit{$D_{temp}$} and \textit{$D_{pol}$}}
\label{tab:maxi4}
\end{table*}
\subsection{Studying the impact of edge density on Reconstruction Error} ~\label{s:exp9} 
 To study the relationship between edge density, $E_d$ and reconstruction error, we calculate the reconstruction error for different edge densities $0.20-0.75$. For this experiment, we consider the \textit{Frob} for Phase-II and $K$ as $7$. Our observations differ with respect to datasets. For example, we observe that the higher the $E_d$, the lower the reconstruction error for $D_{epa}$ as shown in Figure \ref{fig:edge_D1} whereas the reconstruction error is the highest when $E_d$ is $0.2$ and decreases with increase in $E_d$ for $D_{temp}$ as shown in Figure \ref{fig:edge_D2}. We did not observe any trend in $D_{pol}$ and $D_{ws}$ as shown in Figure \ref{fig:edge_D3}. Based on these observations, we conclude there is an optimal $E_d$ for which the reconstruction error is the lowest for a dataset. However, the optimal $E_d$ differs across datasets.

\begin{figure*} [ht]
     \subfigure[]{\label{fig:D3_nei043}\includegraphics[width=1.7in]{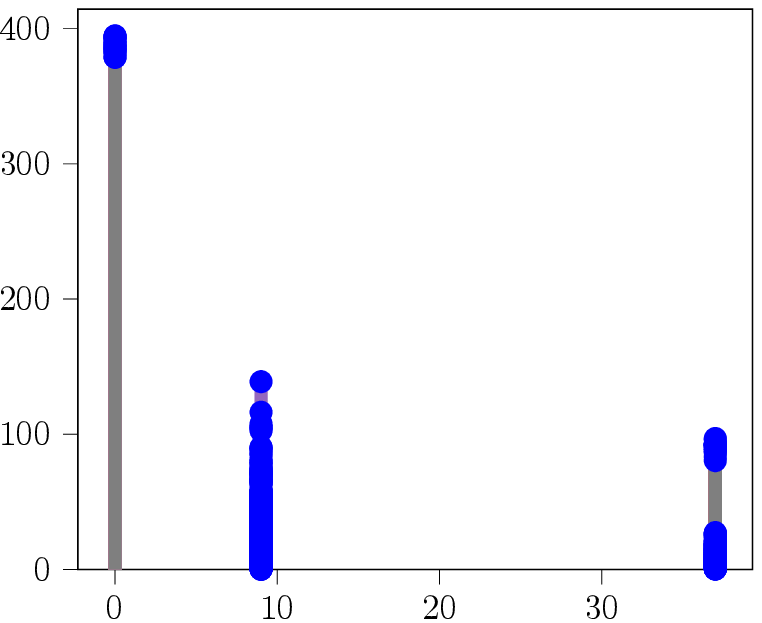}}
    \subfigure[]{ \label{fig:D2_haar040}\includegraphics[width=1.7in]{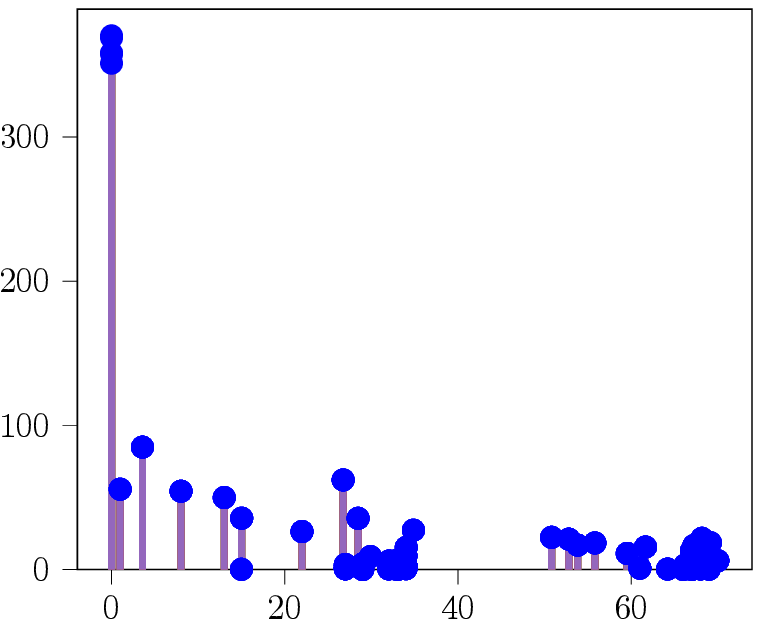}}
    \subfigure[]{\label{fig:D3_dtw061}\includegraphics[width=1.7in]{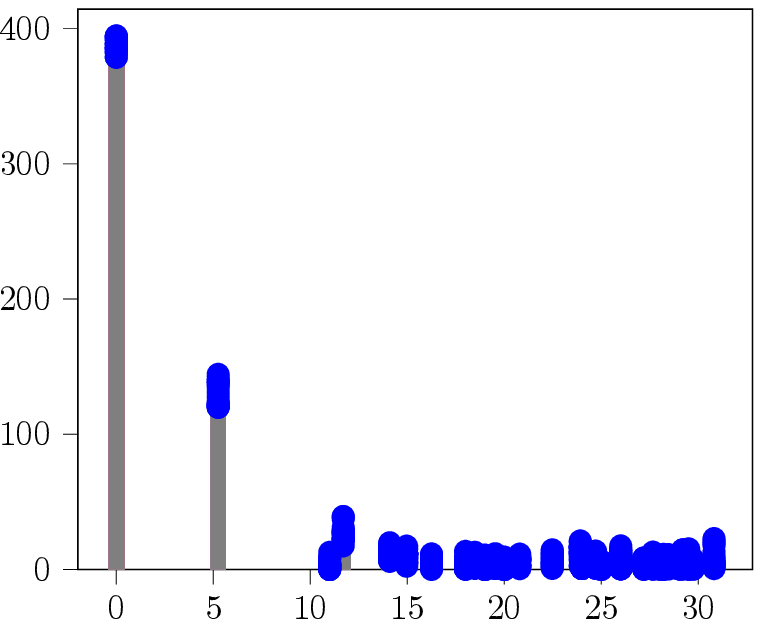}}
     \renewcommand\thefigure{10}
   \caption{ Graph Fourier transforms of $G_{nei}$ with $E_d$ $0.43$ on $D_{pol}$ in Figure 10a, $G_{haar}$ with $E_d$ $0.40$ on $D_{temp}$ in Figure 10b and $G_{dtw}$ with $E_d$ $0.60$ on $D_{pol}$ in Figure 10c is shown}
     \label{fig:GFTs}
\end{figure*}

\subsection{Frequency analysis} ~\label{s:exp8a}

In order to understand the performance of the graph creation approaches discussed in Phase-I, we visualize the Frequency transform of the graphs created given a dataset. As discussed in Section 3.2, Graph Fourier Transform (GFT) is the eigen decomposition of the graph Laplacian, $L$ into eigenvalues, $\Lambda$ and eigenvectors, $V$, i.e., GFT of $L$ is $L=V\Lambda V^{-1}$. Additionally, $GFT$ of the graph signal at the $k-$th time-stamp, $x^k$, is $\hat{x}^k$ which is defined as $\hat{x}^k=V^{-1} x^k$. Therefore, given a dataset, the GFT of the optimal graph topology should comprise of the maximum number of possible distinct eigenvalues which are evenly spread. Additionally, the GFT of the optimal graph topology should be such that the lower the eigenvalues, the higher the amplitudes and vice-versa. Our observations indicate that the GFT of $G_{haar}$ and $G_{dtw}$ ensures optimal graph topology created given a dataset and the $E_d$ whereas $G_{nei}$ has fewer distinct eigenvalues and therefore, is not optimal for $D_{pol}$ and $D_{temp}$. We show some representative examples of our observations in Figure \ref{fig:GFTs}

\subsection{Identifying the Maximum Number of Sampling Sub-sets} ~\label{s:exp11}
In this Subsection, we find the maximum number of sampling sub-sets, $K$ generated by each sampling technique given $\epsilon$. However, most of the sampling techniques proposed in Phase-II except \textit{SRel}, \textit{SMMR} and \textit{SEMMR} only focus on identifying the maximum error given $K$ and therefore, they require $K$ to be pre-specified and can not be modified to identify the maximum number of sampling sub-sets given $\epsilon$. So, we select only \textit{SRel}, \textit{SMMR} and \textit{SEMMR}. We evaluate their performance on $G_{dtw}$, $G_{nei}$, and $G_{haar}$ for $E_d$ as $0.20-0.75$ on $D_{pol}$ and $D_{temp}$ datasets. Our observations indicate that \textit{SMMR} and \textit{SRel} generates the maximum value for $K$ for low and high $E_d$ respectively for a given $\epsilon$.

\begin{figure*} [ht]
     \subfigure[]{\label{fig:casen31}\includegraphics[width=2in]{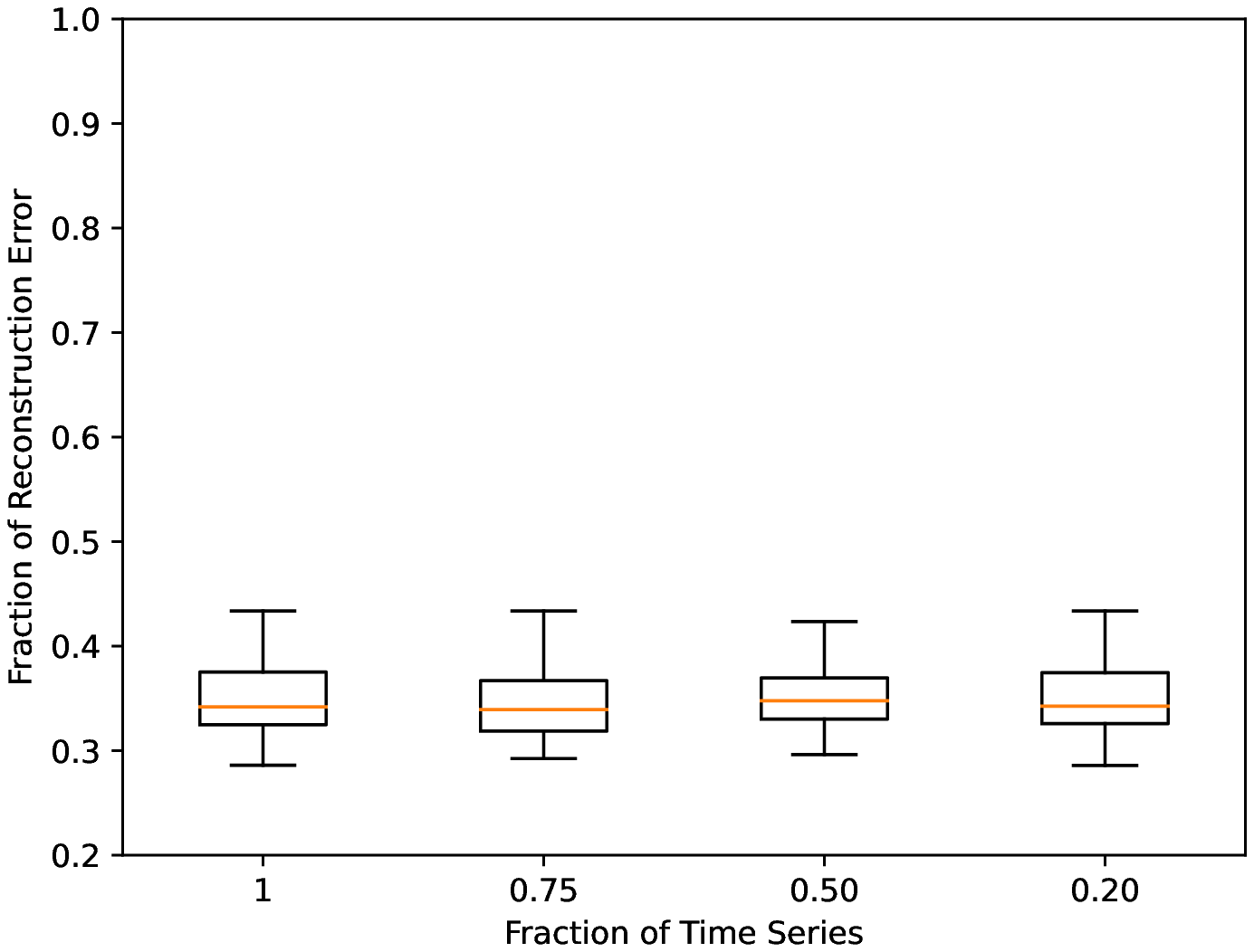}}
    \subfigure[]{ \label{fig:casen32}\includegraphics[width=2in]{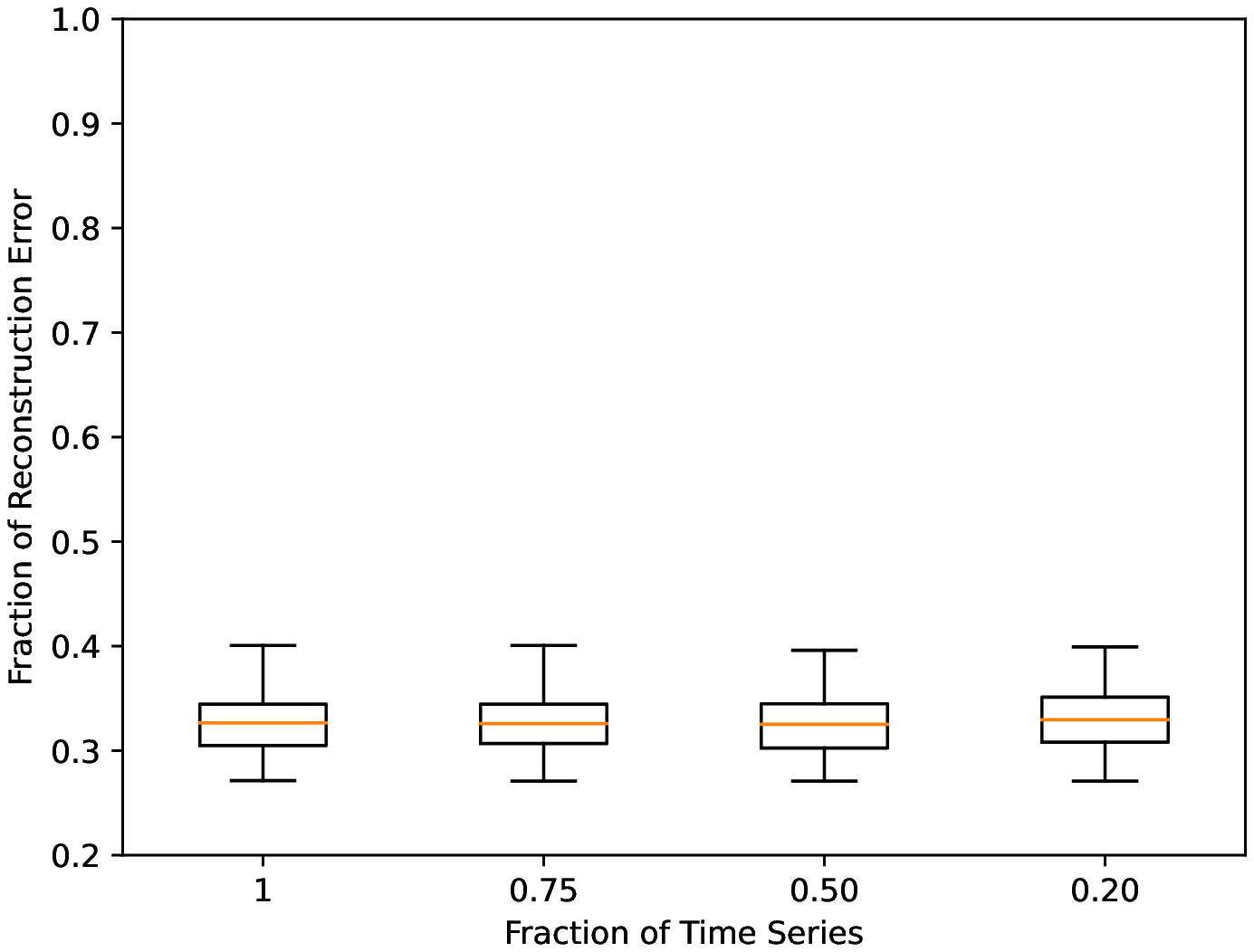}}
    \subfigure[]{\label{fig:casen33}\includegraphics[width=2in]{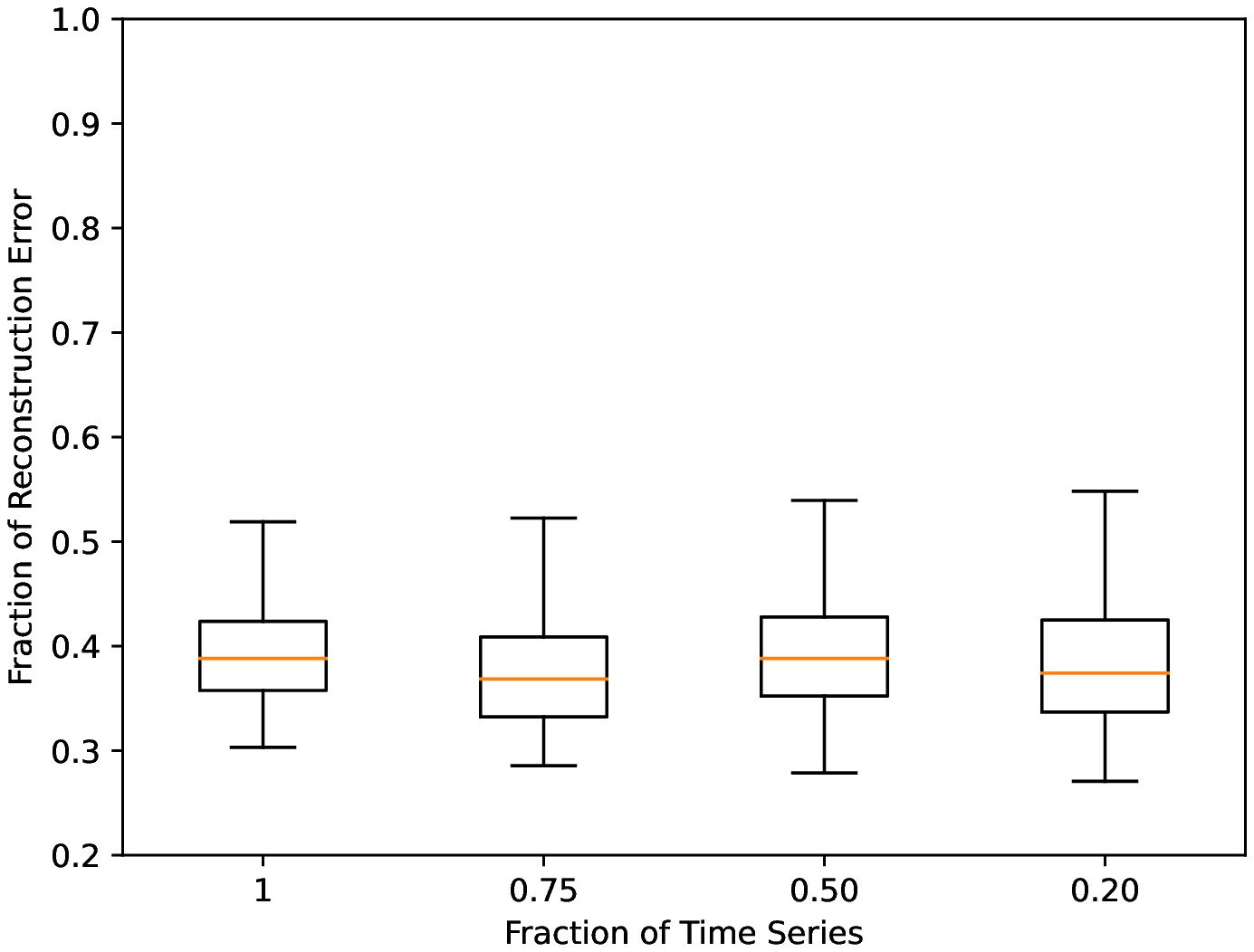}}
     \renewcommand\thefigure{11}
   \caption{ Comparing the reconstruction error for different fraction of time series for \textit{SRel} in Figure 11a, \textit{SMMR} in Figure 11b and \textit{SEMMR} in Figure 11c when $E_d$ is $0.60$ for \textit{$D_{pol}$} is shown. The x-axis represents the fraction of time series and y-axis represents the reconstruction error}
     \label{fig:casestudyn3}
\end{figure*}

\begin{table*} 
 \begin{center}
   \begin{tabular}{|l|l|l|l|l|l|l|l|l|l|l|l|}
\hline
      \textbf{Phase-I} & \textbf{{$E_d$}} & \textbf{{Avg $P_l$}} & \textbf{{Avg $CC$}} & \textbf{TCER} & \textbf{Th} & \textbf{Phase-I} & \textbf{{$E_d$}} & \textbf{{Avg $P_l$}} & \textbf{{Avg  $CC$}} & \textbf{TCER} & \textbf{Th}\\
      \hline
      {\multirow{ 3}{*}{$G_{dtw}$}} &  {{0.23}} & {{inf}} & {{0.91}} & {{0.49}} & {{120}} & {\multirow{ 3}{*}{$G_{haar}$}} &  {{0.23}} & {{2.64}} & {{0.81}} & {{0.97}} & {{60}}\\
      &  {{0.40}} & {{2.26}} & {{0.84}} & {{0.97}} & {{210}} & & {{0.40}} & {{1.77}} & {{0.77}} & {{0.96}} & {{100}}\\
      &  {{0.60}} & {{1.47}} & {{0.89}} & {{0.97}} & {{330}} & &  {{0.61}} & {{1.39}} & {{0.81}} & {{0.93}} & {{185}}\\
      \hline
      {\multirow{ 3}{*}{$G_{nei}$}} &  {{0.23}} & {{1.78}} & {{0.89}} & {{0.89}} & {{5}} & {\multirow{ 3}{*}{$G_{gsp}$}} &  {{0.20}} & {{2.05}} & {{0.39}} & {{0.90}} & {0.01}\\
      
      &  {{0.40}} & {{1.60}} & {{0.86}} & {{0.85}} & {{9}} &  & {{0.40}} & {{1.69}} & {{0.52}} & {{0.89}} & {-4e-06}\\
      
       &  {{0.60}} & {{1.39}} & {{0.84}} & {{0.87}} & {{15}} & &  {{0.59}} & {{1.47}} & {{0.71}} & {{0.89}} & {{-1.1e-05}}\\
      
      \hline
      \end{tabular}
      \end{center}
      \renewcommand\thetable{7}
      \caption{Average path length ($P_l$), clustering co-efficient ($CC$), TCER and the threshold value for $D_{st}$ is shown}
      \label{tab:casestudy}
\end{table*}

\subsection{Evaluation of \textit{SubGraphSample} on partial Time Series} ~\label{s:exp13}
In our previous experiments, we generate $\mathcal{G}$ based on the complete time-series for a dataset. In this Subsection, we analyze the performance of the sampling techniques when the complete time-series is not available. Therefore, in this experiment, we select a fraction, $p$, of the time-series for which to generate similarity graph, $\mathcal{G}^'$, and determine the sampling sets on $\mathcal{G}^'$. We perform this experiment by varying $p$ as $0.80$, $0.50$ and $0.20$ of the time-series. Based on our previous observations, we select $P_{haar}$ in Phase-I and \textit{SRel}, \textit{SMMR} and \textit{SEMMR} in Phase-II. We consider $5$, $7$, $10$ and $13$ sampling sets and the edge densities between $0.20-0.75$. We calculate \textit{average path length}, \textit{clustering co-efficient}, \textit{TCER values} and \textit{reconstruction error}. Our observations indicate that the \textit{average path length}, \textit{clustering co-efficient} are similar for all edge densities irrespective of $p$. Furthermore, we observe that \textit{SRel}, \textit{SMMR} and \textit{SEMMR} yield similar \textit{reconstruction error} irrespective of $p$ for different $E_d$ and the number of sampling sets. We show our observations in Figure \ref{fig:casestudyn3} for $D_{pol}$. 

\section{Conclusions and Future Works}~\label{s:con}
In this paper, we propose \textit{SubGraphSample} which finds the maximum number of \textit{representative sampling subsets} given a \textit{sensor graph}. By finding the maximum number of \textit{representative sampling subsets}, we can alternate querying between these and thus, increase battery longevity significantly. Unlike existing sampling approaches, \textit{SubGraphSample} do not require prior knowledge of the similarity of the sensors and automatically identifies the maximum number of \textit{representative sampling subsets}. We explore $6$ graph creation approaches, propose $2$ new and extend $4$ existing sampling approaches in \textit{SubGraphSample}. However, the suitability and performance of a graph creation approach and sampling approach varies across datasets. Therefore, we propose Algorithm \textit{AutoSubGraphSample} which can autoselect the most suitable approaches given a \textit{sensor graph} and we, further, show the generalizability of \textit{AutoSubGraphSample} given a dataset. We evaluate all possible combination of approaches of \textit{SubGraphSample} on $4$ datasets which shows that the best combination of algorithms can provide $5-13$ times increase in battery life within a $20-40\%$ error bound.

\par As a future work, we will extend \textit{AutoSubGraphSample} to handle multivariate time series and scale to large time series using deep learning-based time series embedding. Furthermore, we aim to merge the current two phases into one in a deep reinforcement learning based model.

\section*{Acknowledgment}
This work has, in part, been supported by the Danish Council for Independent Research (Grant No. 8022-00284B SEMIOTIC).



\bibliographystyle{acm}
\bibliography{stance.bib}

\end{document}